\documentclass[preprint,authoryear,12pt]{elsarticle}
\usepackage{amssymb}
\usepackage{amsfonts}
\usepackage{amsmath}
\usepackage{graphicx}
\usepackage{theorem}
\usepackage{url}

\newcommand{\F}{\mathcal{F}}
\newcommand{\Po}{\mathbb{P}}
\newcommand{\St}{\mathcal{S}}

\newtheorem{prob}{Problem}[section]
\newtheorem{thm}{Theorem}[section]
\newtheorem{prop}{Proposition}[section]
\newtheorem{exm}{Example}[section]
\newtheorem{lem}{Lemma}[section]
\newtheorem{rem}{Remark}[section]
\newproof{pf}{Proof}
\theorembodyfont{\itshape}

\journal{Insurance: Mathematics and Economics}

\begin{document}

\begin{frontmatter}
\title{Explicit Solutions of Optimal Consumption, Investment and Insurance Problems with Regime Switching}

\author[bz]{Bin Zou}
\ead{bzou@ualberta.ca}

\author[bz]{Abel Cadenillas\corref{cor1}}
\ead{abel@ualberta.ca}

\cortext[cor1]{Corresponding author: Central Academic Building 639, Department of Mathematical and Statistical Sciences, University of Alberta, Edmonton, Alberta, Canada T6G 2G1. Telephone:+1-780-492-0572, Fax: +1-780-492-6826.}

\address[bz]{Department of Mathematical and Statistical Sciences, University of Alberta, Canada}

\begin{abstract}
We consider an investor who wants to select her/his optimal consumption, investment and insurance policies.
Motivated by new insurance products, we allow not only the financial market but also the insurable loss to
depend on the regime of the economy.
The objective of the investor is to maximize her/his expected total discounted utility of
consumption over an infinite time horizon.
For the case of hyperbolic absolute risk aversion (HARA) utility functions, we obtain
the first explicit solutions for
simultaneous
optimal consumption, investment, and insurance problems when there is regime switching.
We determine that the optimal insurance contract is either no-insurance
or deductible insurance, and calculate  when it is optimal to buy insurance.
The optimal policy depends strongly on the regime of the economy.
Through an economic analysis, we calculate the advantage of buying insurance.
\vspace{1ex}
\end{abstract}


\end{frontmatter}

\section{Introduction}
In the classical consumption and investment problem,  a risk-averse investor wants to maximize her/his expected discounted utility of consumption by selecting optimal consumption and investment strategies. \citet{merton} was the first to obtain explicit solutions to this problem in continuous time. Many generalizations to Merton's work can be found in  \citet{karatzas1}, \citet{karatzas2}, \citet{sethi}, et c\'{e}tera. In the traditional models for consumption and investment problems, there is only one source of risk that comes from the uncertainty of the stock prices. But in real life, apart from the risk exposure in the financial market, investors often face other random risks, such as property-liability risk and credit default risk. Thus, it is more realistic and practical to extend the traditional models by incorporating an insurable risk. When an investor is subject to an additional insurable risk, buying insurance is a trade-off decision. On  one hand, insurance can provide the investor with compensation and then offset capital losses if the specified risk events occur. On the other hand, the cost of insurance diminishes the investor's ability to consume and therefore reduces the investor's expected utility of consumption.

The initial optimal insurance problem studies an individual who is subject to an insurable risk and seeks the optimal amount of insurance under the utility maximization criterion. Using the expected value principle for premium, \citet{arrow} found the optimal insurance is deductible insurance in discrete time. \citet{promislow} reviewed optimal insurance problems (without investment and consumption). They proposed a general market model and obtained explicit solutions to optimal insurance problems under different
premium principles, such as variance principle, equivalent utility principle, Wang's principle, et c{\'e}tera.

\citet{moore}
combined
Merton's optimal consumption and investment problem and Arrow's optimal insurance problem in continuous time.
They found explicit or numerical solutions for different utility functions (although they did not verify rigourously that the obtained strategies were indeed optimal). \citet{perera} revisited Moore and Young's work by considering their problem in a more general Levy market, and applied the martingale approach to obtain explicit optimal strategies for exponential utility functions. \citet{pirvu} considered the insurable risk to be mortality risk and studied optimal investment, consumption and life insurance problems in a financial market in which the stock price followed a mean-reverting process.

In traditional financial modeling, the market parameters, like the risk-free interest rate, stock returns and volatility, are assumed to be independent of general macroeconomic conditions. However, historical data and empirical research show that the market behavior is affected by long-term economic factors, which may change dramatically as time evolves. Regime switching models use a continuous-time Markov chain with a finite-state space to represent the uncertainty of those long-term economic factors.

\citet{hamilton1} introduced a regime switching model for the first time to capture the movements of the stock prices and showed that the regime switching model represents the stock returns better than the model with deterministic coefficients. Thereafter, regime switching has been applied to model many financial and economic problems. In regard to optimal portfolio selection problems, \citet{zhou} considered a financial market with regime switching and studied the problem under the Markowitz's mean-variance criterion. \citet{sotomayor} considered the problem under the expected utility of consumption maximization criterion, and obtained explicit solutions for HARA utility functions.

In the insurance market, insurance policies can depend on the regime of the economy.
In the case of traditional insurance,
the underwriting cycle has been well documented in the literature.
Indeed, empirical research provides evidence for the dependence of insurance policies' underwriting performance on external economic conditions
(see for instance \citet{grace}, \citet{haley} on property-liability insurance, and \citet{chung} on reinsurance).
In the case of non-traditional insurance, by investigating the comovements of credit default swap (CDS) and the bond/stock markets, \citet{norden} found that CDS spreads are negatively correlated with the price movements of the underlying stocks and such cointegration is affected by the corporate bond volume.

In this paper, we use an observable continuous-time finite-state Markov chain to model the regime of the economy, and allow
both the financial market and the insurance market to depend on the regime. Our objective is to obtain
simultaneously optimal consumption, investment and insurance policies for a risk-averse investor who wants to maximize her/his expected total discounted utility of consumption over an infinite time horizon.
We extend  \citet{sotomayor} by including a random loss in the model and an insurance policy in the control.
The most important difference between the model of \citet{moore} and our paper
 is that they do not allow regime switching, while we allow regime switching in both the financial market and the insurance market.

This paper is organized as follows. In Section \ref{sec_model}, we discribe the problem. The verification theorems are presented in Section \ref{sec_verification}. In Section \ref{sec_explicitsolution}, we obtain explicit solutions for four HARA utility functions. In Section \ref{sec_economicanalysis}, we conduct an economic analysis to investigate the impact of various factors on the optimal policy and calculate the advantage of buying insurance. Section \ref{sec_conclusion} concludes our study.

\section{The Model} \label{sec_model}
Consider a complete probability space $(\Omega,\F,\Po)$ in which a standard Brownian motion $W$ and an observable continuous-time, stationary, finite-state Markov chain $\epsilon$ are defined. Denote by $\St=\{1,2,\cdots,S\}$ the state space of this Markov chain, where $S$ is the number of regimes in the economy. The matrix $Q=(q_{ij})_{S \times S}$ denotes the strongly irreducible generator of $\epsilon$, where $\forall \, i \in \St$, $\sum_{j \in \St} q_{ij}=0$, $q_{ij}>0$ when $j \neq i$ and $q_{ii}=-\sum_{j \neq i}q_{ij}$.

We consider a financial market consisting of two assets,
 a bond with price $P_0$ (riskless asset) and a stock with price $P_1$ (risky asset), respectively. Their prices processes are driven by the following dynamics:
\begin{align*}
dP_0(t) &= r_{\epsilon(t)} P_0(t) dt,\\
dP_1(t) &= P_1(t) (\mu_{\epsilon(t)} dt + \sigma_{\epsilon(t)} dW(t)),
\end{align*}
with initial conditions $P_0(0)=1$ and $P_1(0)>0$. The coefficients $r_i$, $\mu_i$ and $\sigma_i$, $i \in \St$, are all positive constants.

An investor chooses $\pi=\{\pi(t), t\ge 0\}$, the proportion of wealth invested in the stock, and a consumption rate process $c=\{c(t),t\ge 0\}$. We assume the investor is subject to an insurable loss $L(t,\epsilon(t),X(t))$, where $X(t)$ denotes the investor's wealth at time $t$. We shall use the short notation $L_t$ to replace $L(t,\epsilon(t),X(t))$ if there is no confusion.  We use a Poisson process $N$ with intensity $\lambda_{\epsilon(t)}$, where $\lambda_i>0$ for every $i \in \St$, to model the occurrence of this insurable loss. In the insurance market, there are insurance policies available to insure against the loss $L_t$. We further assume the investor can control the payout amount $I(t)$, where $I(t): [0,\infty) \times \Omega \mapsto [0,\infty)$ and $I(t,\omega):=I_t(L(t, \epsilon(t,\omega), X(t,\omega)))$, or in short, $I(t)=I_t(L_t)$. For example, if $\Delta N(t_0)=1$, then at time $t_0$ the investor suffers a loss of amount $L_{t_0}$ but receives a compensation of amount $I_{t_0}(L_{t_0})$ from the insurance policy, so the investor's net loss is $L_{t_0} - I_{t_0}(L_{t_0})$. Following the premium setting used in \citet{moore} (the famous expected value principle), we assume investors pay premium continuously at the rate $P$ given by
\[P(t)= \lambda_{\epsilon(t)} (1+\theta_{\epsilon(t)}) E[I_t(L_t)],\]
where the positive constant $\theta_i,\,i\in\St$, is known as the loading factor in the insurance industry. Such extra positive loading comes from insurance companies' administrative cost, tax, profit, et c{\'e}tera.

Following \citet{sotomayor}, we assume the Brownian motion $W$, the Poisson process $N$ and the Markov chain $\epsilon$ are mutually independent. We also assume that the loss process $L$ is independent of $N$.
We take the $\Po-$augmented filtration $\{ \F_t\}_{t \ge 0}$  generated by $W$, $N$, $L$ and $\epsilon$ as our filtration and define $\F:= \sigma(\cup_{t\ge 0} \F_t)$.

For an investor with triplet strategies $u(t):=(\pi(t),c(t),I(t))$, the associated wealth process $X$ is given by
\begin{equation} \label{wealth}
\begin{split}
dX(t)&= \big( r_{\epsilon(t)}X(t) + (\mu_{\epsilon(t)} - r_{\epsilon(t)}) \pi(t) X(t) -c(t) - \lambda_{\epsilon(t)} (1+\theta_{\epsilon(t)})\\
&\quad \cdot E[I_t(L_t)] \big) dt +\sigma_{\epsilon(t)} \pi(t) X(t) dW(t) - \left( L_t - I_t(L_t)\right)dN(t),
\end{split}
\end{equation}
with initial conditions $X(0)=x>0$ and $\epsilon(0)=i \in \St$.

We define the criterion function $J$ as
\begin{equation}
\label{eqn_criterion} J(x,i;u):= E_{x,i} \left[ \int_0^{+\infty} e^{-\delta t} U(c(t),\epsilon(t)) dt \right] ,
\end{equation}
where $\delta >0$ is the discount rate and $ E_{x,i} $ means conditional expectation given $X(0)=x$ and $\epsilon(0)=i$. We assume that for every $i \in \St$, the utility function $U(\cdot,i)$ is $C^2(0,+\infty)$, strictly increasing and concave, and satisfies the linear growth condition
\[ \exists K>0 \text{ such that }  U(y,i) \le K(1+y), \, \forall \, y>0, i \in \St.\]
Besides, we use the notation $U(0,i):=\lim\limits_{y \to 0^+} U(y,i)$, $\forall \, i \in \St$.

We define the bankruptcy time as
\[ \Theta :=\inf\{ t\ge0: \, X(t) \le 0\}. \]

Since an investor can consume only when her/his wealth is strictly positive, we define
\[ R(\Theta):= \int_\Theta ^ \infty e^{-\delta t} U(c(t),\epsilon(t))dt=\int_\Theta ^ \infty e^{-\delta t} U(0,\epsilon(t))dt.\]

A control $u:=(\pi,c,I)$ is called admissible if $\{u_t\}_{t \ge 0}$ is predictable with respect to the filtration $\{ \F_t\}_{t\ge 0}$ and satisfies, $\forall \, t\ge 0$
\begin{align*}
E_{x,i} &\left[ \int_0^t c(s) ds \right]< +\infty,  \\
E_{x,i} &\left[ \int_0^t \sigma_{\epsilon(s)}^2\pi^2(s) ds \right] < +\infty,  \\
E_{x,i} &\left[ \int_0^\Theta e^{-\delta s} U^+(c(s),\epsilon(s)) ds \right]< +\infty,
\end{align*}
and $I(t) \in \mathcal{I}_t:=\{I: \, 0 \le I(Y) \le Y, \, \text{ where  Y is $\F_t$-measurable} \}.$

We use $\mathcal{A}_{x,i}$ to denote the set of all admissible controls with initial conditions $X(0)=x$ and $\epsilon(0)=i$.
Then we formulate our optimization problem as follows.
\begin{prob} \label{prob}
Select an admissible control $u^*=(\pi^*,c^*,I^*) \in \mathcal{A}_{x,i}$ that maximizes the criterion function $J$. In addition, find the value function
\[ V(x,i) :=\sup_{u\in \mathcal{A}_{x,i}} J(x,i;u). \]
The control $u^*$ is called an optimal control or an optimal policy.
\end{prob}

\cite{moore} also incorporated an insurable risk into the consumption and investment framework. However, they did not consider a regime switching model, or equivalently they assumed that there is only one regime in the economy. Nevertheless, the insurable risk and the coefficients of the financial market most likely depend on the regime of the economy. Hence, in the above regime switching model, we assume that the insurance market (insurable loss and insurance performance) and the financial market are regime dependent. Furthermore, we assume these two markets depend on the same regime.
We mention three examples below.
First, the assumption that the financial market and the insurance market depend on the same regime
is supported by
the bailout case of AIG (see \citet{sjostrom} for details) and the financial derivatives in the insurance industry. Before the crash of the U.S. housing market in 2007, many investors, banks and financial institutions bought obligations constructed from mortgage payments or made loans to the housing agencies. To insure against the credit risk that the obligations or loans may default, they purchased credit default swap (CDS) contracts from insurance companies like AIG. In a CDS contract, the buyer  makes periodic payments to the seller, and in return, receives the par value of the underlying obligation or loan in the event of a default. Apparently, the credit default risk insured by CDS contracts is negatively correlated with the reference entity's stock performance (see \citet{norden} for empirical evidence).
Second, generated by the financial engineering on derivatives, insurance companies have created numerous equity-linked products, such as equity-linked life insurance (see \citet{hardy} for more details on such insurance policy). If the insured of an equity-linked life insurance policy survivals to the expiration, then the beneficiary receives investment benefit that depends upon the market value of the reference equity.
Hence, equity-linked life insurance and its reference equity are affected by the same long-term economic factors.
Third, even in traditional insurance products like property-liability insurance,
there is empirical evidence (see for instance \cite{grace})
that the loading factor $\theta$ depends on the regime of the economy.
Indeed, in those traditional insurance products,
$\lambda_{\epsilon(t)}$ and
$L(t, \epsilon(t,\omega), X(t,\omega))$ might be independent of
$\epsilon(t)$ but $\theta_{\epsilon(t)}$ depends on $\epsilon(t)$.

\section{Verification Theorems}
\label{sec_verification}
Let $\psi: (0,\infty) \times \St \to \mathbb{R}$ be a function with $\psi(\cdot,i) \in C^2(0,\infty), \forall \, i \in \St$. We define the operator $\mathcal{L}_i^u$ by
\[ \mathcal{L}_i^u (\psi) := (r_i x +(\mu_i - r_i) \pi x - c - \lambda_i (1+\theta_i) E[I(L)]) \psi' + \frac{1}{2} \sigma_i^2 \pi^2 x^2 \psi'' - \delta \psi,\]
where $\psi'=\frac{\partial \psi}{\partial x}$ and $\psi''=\frac{\partial^2 \psi}{\partial x^2}$.

\begin{thm} \label{th1}
Suppose $U(0,i)$ is finite, $\forall \, i \in \St$. Let $v(\cdot,i) \in C^2 (0,\infty) $ be an increasing and concave function such that $v(0,i)=\frac{U(0,i)}{\delta}$ for every $i \in \St$.
If $v=v(\cdot,\cdot)$ satisfies the Hamilton-Jacobi-Bellman equation
\begin{align} \label{hjb1}
\sup_{u } \big\{ &\mathcal{L}_i^u v (x,i) + U(c,i) + \lambda_i E[v(x-L+I(L),i) - v(x,i)] \big\} \notag \\
&= - \sum_{j \in \St} q_{ij} \left( v(x,j) - \frac{U(0,j)}{\delta} \right)
\end{align}
for every $ x>0, i \in \St$, and the control $u^*=(\pi^*,c^*,I^*)$ defined by
\begin{equation*}
\begin{split}
u_t^* : = \mathop{\arg\sup}_{u }  \Big( & \mathcal{L}_{\epsilon(t)}^{u}v(X_t^*,\epsilon_t) + U(c,\epsilon_t) \\
& + \lambda_{\epsilon(t)} E[v(X_t^*-L_t+I(L_t),\epsilon_t) - v(X_t^*,\epsilon_t)] \Big) \mathbf{1}_{0 \le t < \Theta}
\end{split}
\end{equation*}
is admissible, then $u^*$ is optimal control to Problem \ref{prob}. In addition, the value function is given by
\[ V(x,i)=v(x,i) + \frac{1}{\delta} E_{x,i} \left[\int_0^\infty e^{-\delta s} dU(0,\epsilon_s) \right],\]
where $dU(0,\epsilon_s):= \sum\limits_{j \in \St} q_{\epsilon_s,j} U(0,j) ds$.

Furthermore, if the utility function does not depend on the regime, namely $U(y,i)=U(y)$, for every $i \in \St$, then the value function $V(x,i)=v(x,i)$.
\end{thm}

\begin{pf}
$\forall \, u \in \mathcal{A}_{x,i}$, consider $f(t,X_t,\epsilon_t):=e^{-\delta t} (v(X_t,\epsilon_t) - \frac{U(0,\epsilon_t)}{\delta})$. By applying Ito's formula for Markov-modulated processes (see, for instance, \citet{sotomayor}), we get
\begin{align} \label{fproc}
f(t,X_t,\epsilon_t) &=   \int_0^t e^{-\delta s} \Big( \mathcal{L}_{\epsilon(s)}^{u(s)} v(X_s,\epsilon_s)  + \lambda_{\epsilon(s)} \left[v(X_s-L_s+I_s,\epsilon_s)-v(X_s,\epsilon_s) \right]\notag\\
&\quad +\sum_{j \in \St} q_{\epsilon_s,j} \big(v(X_s,j)-\frac{U(0,j)}{\delta} \big) + U(0,\epsilon_s)\Big) ds \notag \\
&\quad +v(X_0,\epsilon_0)-\frac{U(0,\epsilon_0)}{\delta} +  m_t^f,
\end{align}
where $\{m_t^f\}_{t\ge 0}$ is a $\Po-$martingale with $m_0^f=0$.

Let $0<a<X_0=x<b<\infty$ and define a stopping time $\tau:=\inf\{ t\ge 0: X_t \le a \text{ or } X_t \ge b \}$. Then by replacing $t$ by $t\wedge \tau$ in \eqref{fproc}, taking conditional expectation and applying the HJB equation \eqref{hjb1}, we obtain
\begin{align*}
E_{x,i} [f(t\wedge \tau,X_{t\wedge \tau},\epsilon_{t\wedge \tau})] &\le -E_{x,i} [\int_0^{t\wedge \tau} e^{-\delta s} (U(c_s,\epsilon_s)-U(0,\epsilon_s)) ds] \\
&\quad +v(x,i)-\frac{U(0,i)}{\delta}.
\end{align*}

Let $a \downarrow 0$, $b \uparrow +\infty$ and $t \to \infty$. Then $t \wedge \tau \to \Theta$. Since $f$ is continuous, we obtain
\[ f(t\wedge \tau,X_{t\wedge \tau},\epsilon_{t\wedge \tau}) \to f(\Theta,0,\epsilon_{\Theta})=0, \text{ when }a \downarrow 0, b \uparrow +\infty, t \to \infty.\]
Hence,
\begin{equation} \label{ineq1}
v(x,i)-\frac{U(0,i)}{\delta}-E_{x,i} \left[\int_0^{\Theta} e^{-\delta s} (U(c_s,\epsilon_s)-U(0,\epsilon_s)) ds\right] \ge 0.
\end{equation}

Define $g(t,\epsilon_t):=-e^{-\delta t} \frac{U(0,\epsilon_t)}{\delta}$. Applying Ito's formula to $g(t,\epsilon_t)$ yields
\[ g(t,\epsilon_t)-g(0,\epsilon_0) = \int_0^t  e^{-\delta s} \left( U(0,\epsilon_s) - \frac{1}{\delta} \sum_{j \in \St} q_{\epsilon_s,j} U(0,j) \right)ds +m_t^g,\]
where $\{m_t^g\}_{t \ge 0}$ is a square-integrable martingale with $m_0^g=0$.

Taking conditional expectation and applying the monotone convergence theorem to the above equality, we get
\begin{equation*}
\frac{U(0,i)}{\delta}=E_{x,i} \left[ \int_0 ^ \infty e^{-\delta s} U(0,\epsilon_s)ds \right]
- \frac{1}{\delta}E_{x,i}\left[\int_0^\infty e^{-\delta s} dU(0,\epsilon_s) \right],
\end{equation*}
and then
\begin{align*}
&v(x,i)-\frac{U(0,i)}{\delta}-E_{x,i} \left[\int_0^{\Theta} e^{-\delta s} (U(c_s,\epsilon_s)-U(0,\epsilon_s)) ds\right]\\
=&v(x,i) - E_{x,i} \left[ \int_0 ^ \infty e^{-\delta s} U(0,\epsilon_s)ds -\frac{1}{\delta}\int_0^\infty e^{-\delta s} dU(0,\epsilon_s)\right]\\
&-E_{x,i} \left[\int_0^{\Theta} e^{-\delta s} (U(c_s,\epsilon_s)-U(0,\epsilon_s)) ds\right]\\
=&v(x,i)+\frac{1}{\delta}E_{x,i}\left[\int_0^\infty e^{-\delta s} dU(0,\epsilon_s) \right]-E_{x,i}\left[\int_0^{\Theta} e^{-\delta s}  U(c_s,\epsilon_s)ds ds\right]\\
&-E_{x,i}\left[\int_\Theta^\infty e^{-\delta s} U(0,\epsilon_s)) ds\right].
\end{align*}

Hence, the inequality \eqref{ineq1} can be rearranged as
\[(x,i) + \frac{1}{\delta} E_{x,i} \left[\int_0^\infty e^{-\delta s} dU(0,\epsilon_s) \right] \ge E_{x,i}\left[ \int_0^\infty e^{-\delta s} U(c_s,\epsilon_s)ds \right] = J(x,i;u),\]
and the equality will be achieved when $u=u^*$.

If the utility function does not depend on the regime, then $dU(0,\epsilon_s)=U(0) \cdot \sum\limits_{j\in \St} q_{ij}ds =0$, and so $V(x,i)=v(x,i)$.
$\hfill \Box$
\end{pf}

$U(\cdot,i)$ is an increasing function for every $i \in \St$, so if $U(0,i)$ is not finite, then $U(0,i) = - \infty$. The following theorem deals with the case when $U(0,i) = - \infty$, $\forall \, i \in \St$.

\begin{thm} \label{th2}
Suppose $U(0,i)=-\infty$ for every $i \in \St$. Let $v(\cdot,i) \in C^2 (0,\infty) $ be an increasing and concave function such that $v(0,i)=-\infty$ for every $i \in \St$.
If $v=v(\cdot,\cdot)$ satisfies the Hamilton-Jacobi-Bellman equation
\begin{align} \label{hjb2}
\sup_{u } \big\{ &\mathcal{L}_i^u v (x,i) + U(c,i) + \lambda_i E[v(x-L+I(L),i) - v(x,i)] \big\}\notag \\
&= - \sum_{j \in \St} q_{ij} v(x,j)
\end{align}
for every $ x>0, i \in \St$, and the control $u^*=(\pi^*,c^*,I^*)$ defined by
\begin{equation*}
\begin{split}
u_t^* : = \mathop{\arg\sup}_{u }  \Big( &\mathcal{L}_{\epsilon(t)}^{u}v(X_t^*,\epsilon_t) + U(c,\epsilon_t) \\
& + \lambda_{\epsilon(t)} E[v(X_t^*-L_t+I(L_t),\epsilon_t) - v(X_t^*,\epsilon_t)] \Big) \mathbf{1}_{0 \le t < \Theta}
\end{split}
\end{equation*}
is admissible, then $u^*$ is an optimal control to Problem \ref{prob} and the value function is $V(x,i)=v(x,i)$.
\end{thm}

\begin{pf}
Define $h(t,X_t,\epsilon_t):=e^{-\delta t} v(X_t,\epsilon_t)$. For any admissible control $u$, by following a similar argument as in Theorem \ref{th1}, we obtain
\begin{align*}
E_{x,i}\left[h(t\wedge \tau, X_{t\wedge \tau},\epsilon_{t\wedge \tau})\right]&=E_{x,i}\Big[ \int_0^{t\wedge \tau} e^{-\delta s} \Big( \mathcal{L}_{\epsilon(s)}^{u(s)} v(X_s,\epsilon_s)+ \sum_{j\in \St} q_{\epsilon_s,j}v(X_s,j) \\
&\; +\lambda_i E[v(X_s-L_s+I_s,\epsilon_s) - v(X_s,\epsilon_s)]  \Big) ds \Big]+v(x,i)\\
&\le v(x,i)-E_{x,i}\left[ \int_0^{t\wedge \tau} e^{-\delta s} U(c_s,\epsilon_s) ds \right].
\end{align*}
$E_{x,i} \left[ \int_0^{t\wedge \tau} e^{-\delta s} U(c(s),\epsilon(s)) ds\right]$ is well defined and finite, because $u$ is an admissible control and $U$ satisfies the linear growth condition.
Then the above inequality becomes
\begin{equation*}
v(x,i)\ge E_{x,i}\left[h(t\wedge \tau, X_{t\wedge \tau},\epsilon_{t\wedge \tau})\right] + E_{x,i}\left[ \int_0^{t\wedge \tau} e^{-\delta s} U(c_s,\epsilon_s) ds \right].
\end{equation*}

By assumption, $v(\cdot,i)$ is increasing in $(0,\infty)$ and $v(0,i)=\frac{U(0,i)}{\delta}=-\infty$ for every $i \in \St$, so
\[E_{x,i}[h(t\wedge \tau, X_{t\wedge \tau},\epsilon_{t\wedge \tau})] \ge E_{x,i} \left[ \int_{t\wedge \tau}^\infty e^{-\delta s} U(0,\epsilon_s) ds \right].\]

By letting $a\downarrow 0$, $b\uparrow +\infty$ and $t \to \infty$, and applying the monotone convergence theorem, we obtain
\[ v(x,i)\ge E_{x,i}\left[ \int_0^\Theta e^{-\delta s} U(c_s,\epsilon_s) ds\right] +E_{x,i}\left[ \int_\Theta^\infty e^{-\delta s} U(0,\epsilon_s) ds\right]=J(x,i;u),\]
and the equality holds when $u=u^*$.  $\hfill \Box$
\end{pf}

\section{Explicit Solutions of Value Function and Optimal Strategies}
\label{sec_explicitsolution}
In this section, we obtain explicit solutions to optimal consumption, investment and insurance problem when there is regime switching in the economy. We assume the utility function is of HARA type and the insurable loss $L$ is proportional to the investor's wealth, $L(t,\epsilon(t),X(t))=\eta_{\epsilon(t)} \, l_t \,X_t$. Here for every $i \in \St$, $\eta_i >0$ measures the intensity of the insurable loss in regime $i$, and for every $t \ge 0$, $l_t$ denotes the loss proportion at time $t$. We assume that $l_t$ is $\F_t-$measurable and $l_t \in (0,1)$ for all $t \ge 0$.

To obtain optimal policy, we first construct a candidate policy at time $t$, which is a function of $(x,i,l)$, namely, $\pi^*=\pi^*(x,i,l)$, $c^*=c^*(x,i,l)$ and $I^*=I^*(x,i,l)$ (In fact, we find $\pi^*$ and $c^*$ are independent of $l$). The candidate policy is indeed optimal once we can prove it is an admissible policy.

We rewrite the HJB equation \eqref{hjb1} as
\begin{align} \label{HJB1}
&\quad\max_{\pi} \Big[ (\mu_i - r_i) \pi x v'(x,i) + \frac{1}{2} \sigma_i^2 \pi^2 x^2 v''(x,i) \Big] + \max_{c} \Big[U(c,i)-cv'(x,i) \Big] \notag \\
&\quad+ \lambda_i \max_{I} \Big[ Ev(x-\eta_i lx+I(\eta_i lx),i)-(1+\theta_i) E(I(\eta_i lx)) v'(x,i) \Big] \notag \\
&=(\delta + \lambda_i) v(x,i)-r_i x v'(x,i)  - \sum_{j \in \St} q_{ij} \left( v(x,j) - \frac{U(0,j)}{\delta} \right),
\end{align}
and the HJB equation \eqref{hjb2} as
\begin{align} \label{HJB2}
&\quad\max_{\pi} \Big[ (\mu_i - r_i) \pi x v'(x,i) + \frac{1}{2} \sigma_i^2 \pi^2 x^2 v''(x,i) \Big] + \max_{c} \Big[U(c,i)-cv'(x,i) \Big] \notag \\
&\quad+ \lambda_i \max_{I} \Big[ Ev(x-\eta_i lx+I(\eta_i lx),i)-(1+\theta_i) E(I(\eta_i lx)) v'(x,i) \Big] \notag \\
&=(\delta + \lambda_i) v(x,i)-r_i x v'(x,i)  - \sum_{j \in \St} q_{ij} v(x,j).
\end{align}

We conjecture that $v(\cdot,i)$ is strictly increasing and concave for every $i \in \St$. Then a candidate for $\pi^*$ is given by
\begin{equation} \label{candidateofpi}
\pi^*(x,i) =  - \frac{(\mu_i-r_i) v'(x,i)}{\sigma_i^2 x v''(x,i)}.
\end{equation}

Since $U'$ is strictly decreasing, the inverse of $U'$ exists. Then a candidate for $c^*$ is given by
\begin{equation} \label{candidateofc}
c^*(x,i)=(U')^{-1}(v'(x,i),i).
\end{equation}

For the optimal insurance, we have the following Lemma and Theorem.

\begin{lem}\label{lem}
$\forall \, x>0 \text{ and }i \in \St$, denote $z_0:=\eta_i l_0 x$, where constant $l_0 \in (0,1)$. We denote the optimal insurance policy by $I^*$. Then we have
\begin{description}
\item (a) $I^*(x,i,l_0)=0$  if and only if
\begin{equation*}
(1+\theta_i) v'(x,i) \ge v'(x-z_0,i).
\end{equation*}
\item (b) $0<I^*(x,i,l_0)<z_0$ if and only if
\begin{equation*}
(1+\theta_i) v'(x,i) = v'(x-z_0+I^*(x,i;l_0),i).
\end{equation*}
\end{description}
\end{lem}

\begin{pf}
$\forall \, i \in \St$, we use the notation $z:=\eta_i l x$. We then break the proof into four steps.

\emph{Step 1:} We want to show that $I^*(x,i,l) \neq z$, $\forall \, l \in (0,1)$.

Assume to the contrary that $\exists \, l_0 \in (0,1)$ such that $I^*(x,i,l_0)=I^*(z_0)=z_0$. Consider $\bar{I}(x,i,l):=I^*(x,i,l)-\zeta G(l)$, where $\zeta>0$ and $G(l)=1$ when $l_0 - \rho < l \le l_0+\rho$ and $0$ otherwise, $\rho>0$. Here we choose small $\zeta$ and $\rho$ to ensure that $0 \le \bar{I}(z) \le z$. Let
\[f^I(x,i,l;I):=E \left[v(x-z+I(z),i)\right]-(1+\theta_i) E \left[I(z)\right] v'(x,i).\]
Since $I^*$ is the maximizer of $f^I(x,i,l;I)$, we have
\begin{equation*}
 E\big[ v(x-z + \bar{I}(z), i) -v(x-z + I^*(z), i) \big]
\le (1+\theta_i) E\big[ \bar{I}(z) - I^*(z) \big] v'(x,i).
\end{equation*}

Using Taylor expansion and letting $\zeta \to 0^+$, we get
\[ (1+\theta_i) E\big[ G(l)] v'(x,i) \le E \big[ v'(x-z + I^*(z), i) G(l) \big] .\]

Letting $\rho \to 0^+$ ($z \to z_0$) and applying the mean value theorem of integrals, we obtain
\[ (1+\theta_i) v'(x,i) \le  v'(x-z_0 + I^*(z_0), i) =v'(x,i), \]
which is a contradiction since $v'(x,i)>0$ and $\theta_i>0$, $\forall i \in \St$.

\emph{Step 2:} We want to show that $I^*(x,i,l_0)=0 \Rightarrow (1+\theta_i) v'(x,i) \ge v'(x-z_0,i)$.

To this purpose, we consider $\bar{I}'(x,i,l):=I^*(x,i,l)+\zeta G(l)$. For small enough $\zeta$ and $\rho$, we have $0 \le \bar{I}'(z) \le z$. Then a similar argument as above gives the desired result \[(1+\theta_i)v'(x,i) \ge v'(x-z_0 + I^*(x,i,l_0), i) =v'(x-z_0,i).\]

\emph{Step 3:} We want to show that $0<I^*(x,i,l_0)<z_0 \Rightarrow (1+\theta_i) v'(x,i) = v'(x-z_0+I^*(x,i,l_0),i)$.

In this step, we consider $\bar{I}(x,i,l)$ and $\bar{I}'(x,i,l)$. From the results in \emph{Step 1} and \emph{Step 2}, we obtain $(1+\theta_i) v'(x,i) \le  v'(x-z_0 + I^*(x,i,l_0), i)$ and $(1+\theta_i)v'(x,i) \ge v'(x-z_0 + I^*(x,i,l_0), i)$ at the same time, and thus the equality is achieved.

\emph{Step 4:} We want to show that $(1+\theta_i) v'(x,i) \ge v'(x-z_0,i) \Rightarrow I^*(x,i,l_0)=0$.

We assume to the contrary that $I^*(x,i,l_0)>0$. Then the results above give $(1+\theta_i) v'(x,i) = v'(x-z_0+I^*(x,i,l_0),i)<v'(x-z_0,i)$, which is a contradiction to the given condition. A similar method also applies to the proof of $(1+\theta_i) v'(x,i) = v'(x-z_0+I^*(x,i,l_0),i) \Rightarrow 0<I^*(x,i,l_0)<z_0$. $\hfill \Box$
\end{pf}

\begin{thm} \label{opin}
The optimal insurance is either no insurance or deductible insurance (almost surely).
\begin{description}
\item(a) The optimal insurance is no insurance $I^*(x,i,l)=0, \forall i \, \in \St$, when
\begin{equation} \label{con1}
(1+\theta_i) v'(x,i) \ge v'\big((1 - \eta_i \, {\rm ess \, sup} ( l)) \, x,i \big).
\end{equation}
\item(b) The optimal insurance is deductible insurance $I^*(x,i,l) = (\eta_i l x - d_i)^+$, $\forall \, i \, \in \St$, when there exists $d_i:=d_i(x) \, \in (0,x)$ satisfying
\begin{equation} \label{con2}
(1+\theta_i) v'(x,i) = v'(x - d_i,i).
\end{equation}
\end{description}
\end{thm}

\begin{pf} We complete the proof in three steps.

\emph{Step 1:} We want to prove Case (a).

Assume there exists $l_0 \in (0,1)$ such that $0<I^*(x,i,l_0)<z_0$. Then according to (b) in Lemma \ref{lem}, we have $(1+\theta_i) v'(x,i) = v'(x-z_0+I^*(x,i,l_0),i)$. Define the set $N_l:=\{\omega\in \Omega: l(\omega) >  {\rm ess \, sup} ( l)\}$ (we have $\Po \{N_l\}=0$). If $l_0 \le {\rm ess \, sup} ( l)$ (on the set $N_l^c$), then $v'((1 - \eta_i \, {\rm ess \, sup} ( l)) \, x,i )>v'(x-z_0+I^*(x,i,l_0),i)=(1+\theta_i) v'(x,i)$, which is a contradiction to the given condition. Therefore $I^*(x,i,l)=0$ on the set $N_l^c$.

Besides, if two policies $I_1$ and $I_2$ only differ on a negligible set, we have $f^I(i,l;I_1)=f^I(i,l;I_2)$, because the integration of a bounded function on a negligible set is zero.

\emph{Step 2:} We want to prove Case (b).

We notice that $v'(\cdot,i)$ is a strictly decreasing function, so if such $d_i$ exists, it must be unique. We then break our discussion into two disjoint scenarios.

(i) $0<l_0 \le \dfrac{d_i}{\eta_i x}  $

In this scenario, $\eta_i l_0 x \le d_i$, so we have
\[v'((1-\eta_i l_0)x,i) \le v'(x-d_i,i)=(1+\theta_i)v'(x,i).\]
Then by part (a) of Lemma \ref{lem}, we obtain
\[I^*(x,i,l_0)=0=(\eta_i l_0 x -d_i)^+.\]

(ii) $\dfrac{d_i}{\eta_i x}< l_0 <1$

In this scenario, we have $0<I^*(x,i,l_0)<z_0$ since $v'((1-\eta_i l_0)x,i) > v'(x-d_i,i)=(1+\theta_i)v'(x,i)$. Then the result in Lemma \ref{lem} shall give \[(1+\theta_i)v'(x,i)=v'(x-z_0+I^*(x,i,l_0),i)=v'(x-d_i,i).\]
Due to the monotonicity of $v'(\cdot,i)$, we must have
\[I^*(x,i,l_0)=\eta_i l_0 x-d_i=(\eta_i l_0 x-d_i)^+.\]

\emph{Step 3:} We want to show that either \eqref{con1} or \eqref{con2} holds.

If condition \eqref{con1} fails, then
\[v'(x,i) < (1+\theta_i) v'(x,i) < v'((1-\eta_i \, {\rm ess \, sup} ( l)) \, x,i ) \le v'(0,i),\]
where $v'(0,i):=\lim_{x \to 0} v'(x,i)$. Since $v'(\cdot,i)$ is continuous and strictly decreasing in $[0,x]$, there must exist a unique $d_i \in (0,x)$ such that \[(1+\theta_i)v'(x,i)=v'(x-d_i,i).\]
If \eqref{con2} has no solution in $(0,x)$, then
\[(1+\theta_i)v'(x,i) \ge v'(0,i) \ge v'((1-\eta_i \, {\rm ess \, sup} ( l)) \, x,i ).\]

Therefore, we conclude that the optimal insurance is either no insurance or deductible insurance. $\hfill \Box$
\end{pf}

\begin{rem}
The optimal insurance $I^*$ also satisfies the usual properties: $I_t^*(0)=0$ and $I_t^*(\cdot)$ is an increasing function of the loss.
\end{rem}

To find explicit solutions to the optimal consumption, investment and insurance problem, we consider four utility functions of HARA class. The first three utility functions do not depend on the market regimes:
\begin{itemize}
\item[1.] $U(y,i)=\ln (y) , \, y>0$,
\item[2.] $U(y,i)=-y^\alpha,\,y>0,\,\alpha<0$,
\item[3.] $U(y,i)=y^\alpha,\,y>0,\,0<\alpha<1$.
\end{itemize}
The fourth utility function depends on the regime of the economy and we assume there are two regimes in the economy ($S=2$).
\begin{itemize}
\item[4.] $U(y,i)=\beta_i y^{1/2}, \, y>0,\, \beta_i>0,i=1,2.$
\end{itemize}

All these four utility functions are $C^2(0,\infty)$, strictly increasing and concave, and satisfy the linear growth condition. To be specific, we can take $K=1$ for the first three utility functions and $K=\max\{\beta_1,\beta_2\}$ for the last one.

\subsection{$U(y,i)=\ln (y) , y>0, \forall \, i \in \St$}
\label{subsec_log}
In this case, a solution to the HJB equation \eqref{HJB2} is given by
\begin{equation}
\hat{v}(x,i)=\frac{1}{\delta} \ln(\delta x) + \hat{A}_i, \, i \in \St,
\end{equation}
where the constants $\hat{A}_i$, $i\in \St$, will be determined below.

Since $\hat{v}'(x,i)=\frac{1}{\delta x}$, $\hat{v}''(x,i)=-\frac{1}{\delta x^2}$ and $(U')^{-1}(y,i)=\frac{1}{y}$, we obtain from \eqref{candidateofpi} and \eqref{candidateofc} that
\[ \pi^*(x,i)=\frac{\mu_i - r_i}{\sigma_i^2} \quad  \text{and} \quad c^*(x,i)=\delta x.\]

Solving $(1+\theta_i)\hat{v}'(x,i)=\hat{v}'(x-d_i,i)$ gives $d_i=\frac{\theta_i}{1+\theta_i}x \in (0,x)$. Then by Theorem \ref{opin},
\[I^*(x,i,l)=\left(\eta_i l-\frac{\theta_i}{1+\theta_i} \right)^+x.\]

Therefore, the HJB equation \eqref{HJB2} reads as
\[ \frac{r_i}{\delta} +\frac{\gamma_i}{\delta}+\frac{\lambda_i}{\delta} \hat{\Lambda}_i-1=\delta \hat{A}_i - \sum_{j\in \St} q_{ij} \hat{A}_j,\]
where $\gamma_i:=\frac{1}{2} \frac{(\mu_i - r_i)^2}{\sigma^2_i}$ and
\[\hat{\Lambda}_i:= E\left[\ln \left(1-\eta_i l+ \left(\eta_i l-\frac{\theta_i}{1+\theta_i}\right)^+ \right)\right] - (1+\theta_i) E \left[ \left(\eta_i l-\frac{\theta_i}{1+\theta_i}\right)^+ \right].\]

Let 
$\vec{\hat{A}}=(\hat{A}_1,\hat{A}_2,\cdots,\hat{A}_S)'$, 
$\vec{r}=(r_1,r_2,\cdots,r_S)'$, $\vec{\gamma}=(\gamma_1,\gamma_2,\cdots,\gamma_S)'$, $\vec{\lambda \hat{\Lambda}}=(\lambda_1 \hat{\Lambda}_1,\lambda_2 \hat{\Lambda}_2,\cdots,\lambda_S \hat{\Lambda}_S)'$, $\mathbf{1}=(1,1,\cdots,1)'_{S\times 1}$ and $\mathbb{I}$ be the $S \times S$ identity matrix. Then the constant vector $\vec{\hat{A}}$
satisfies the linear system
\begin{equation} \label{linear1}
 (\delta \mathbb{I} - Q) \vec{\hat{A}} = \frac{1}{\delta} \left( \vec{r}+\vec{\gamma}+\vec{\lambda \hat{\Lambda}} -\delta \mathbf{1} \right).
\end{equation}

\begin{prop}
The function $\hat{v}=\hat{v}(\cdot,\cdot)$, given by
\[ \hat{v}(x,i)=\begin{cases}
\frac{1}{\delta} \ln(\delta x) + \hat{A}_i, & x>0\\
-\infty, &x=0 \end{cases} \]
where $\vec{\hat{A}}=(\hat{A}_1,\hat{A}_2,\cdots,\hat{A}_S)'$ solves the linear system \eqref{linear1}, is the value function of Problem \ref{prob}. Furthermore, the policy given by
\[ u^*(t)=(\pi^*(t),c^*(t),I^*(t))=\left(\frac{\mu_{\epsilon(t)}-r_{\epsilon(t)}}{\sigma_{\epsilon(t)}^2},\delta X_t^*,
\left(\eta_{\epsilon(t)} \, l_t-\frac{\theta_{\epsilon(t)}}{1+\theta_{\epsilon(t)}} \right)^+ X_t^* \right) \]
is optimal policy of Problem \ref{prob}.
\end{prop}

\begin{pf}
The function $\hat{v}(\cdot,i)$ defined above is a smooth function which is strictly increasing and concave such that $\hat{v}(0,i)=-\infty$, for every $i \in \St$. By the construction of the vector $\vec{\hat{A}}$, $\hat{v}$ satisfies the HJB equation \eqref{HJB2}.

To show that the candidate policy is admissible, we consider an upper bound process $Z$ of $X^*$
\[ \frac{dZ_t}{Z_t}= \left(r_{\epsilon(t)} - \delta + 2\gamma_{\epsilon(t)} \right)dt+\frac{\mu_{\epsilon(t)}-r_{\epsilon(t)}}{\sigma_{\epsilon(t)}}dW(t),\]
with initial value $Z(0)=x$.

Solving the above SDE gives
\[ Z_t = x \exp \left\{ \int_0^t \left(r_{\epsilon(s)} - \delta + \gamma_{\epsilon(s)} \right) ds + \int_0^t \frac{\mu_{\epsilon(s)}-r_{\epsilon(s)}}{\sigma_{\epsilon(s)}} dW(s)\right\}.\]
By the definition of $Z$, we have $  X_t^* \le Z_t$, $\forall t \ge 0$. Notice that if $\epsilon(t)=i$ for $t\in (t_1,t_2]$, then
\[\int_{t_1}^{t_2} \frac{\mu_{\epsilon(s)}-r_{\epsilon(s)}}{\sigma_{\epsilon(s)}} dW(s)=\frac{\mu_i-r_i}{\sigma_i} (W(t_2)-W(t_1)).\]
So
$\int_0^t \frac{\mu_{\epsilon(s)}-r_{\epsilon(s)}}{\sigma_{\epsilon(s)}} dW(s)$ is a linear combination of independent Brownian motions. By the exponential martingale property of a Brownian motion, we have
\[ E \left[ \exp\left( \int_0^t \frac{\mu_{\epsilon(s)}-r_{\epsilon(s)}}{\sigma_{\epsilon(s)}} dW(s)\right) \right]= \exp\left(\int_0^t \gamma_{\epsilon(s)} ds\right).\]

For the candidate of optimal investment proportion $\pi^*$,
\begin{equation*}
E_{x,i} \left[ \int_0^t \sigma_{\epsilon(t)}^2 (\pi^*(s))^2 ds \right] \le 2\, \gamma_M \, t < \infty, \, \forall t\ge 0,
\end{equation*}
where $\gamma_M=\max\limits_{i \in \St} \{\gamma_i\}$.

Since $c^*(t)=0, \forall \, t>\Theta$, for the candidate of optimal consumption $c^*$,
\begin{align*}
E_{x,i} \left[ \int_0^t c^*(s)ds \right] &=E_{x,i} \left[ \int_0^t c^*(s) 1_{s \le \Theta}ds \right]  \\
&\le \delta E_{x,i} \left[ \int_0^t X_s^* ds \right] \le \delta E_{x,i} \left[ \int_0^t Z_s ds \right]\\
&\le \delta x   \int_0^t e^{K_1 s} ds= \frac{\delta x}{K_1} \left(e^{K_1\,t}-1 \right) < \infty, \, \forall t\ge 0,
\end{align*}
where $K_1=\max\limits_{i \in \St} \{r_i-\delta+2\gamma_i\}$.

For the candidate of optimal insurance $I^*$, $\forall \, \F_t$-measurable random variable $Y$,
$0\le I^*_t(Y)= \Big(Y -\frac{\theta_{\epsilon(t)}}{1+\theta_{\epsilon(t)}} X_t^* \Big)^+ \le Y$, so $I^*_t \in \mathcal{I}_t$.

Furthermore, we have
\begin{equation*}
\begin{split}
\quad E_{x,i} \left[ \int_0^\Theta e^{-\delta s} \ln^+ (c_s^*) ds \right] &\le E_{x,i} \left[ \int_0^\infty e^{-\delta s} |\ln (\delta Z_s)| ds \right] \\
&\le \frac{1}{\delta}|\ln(\delta x)| + K'_1 \int_0^\infty e^{-\delta s} s ds\\
&\quad + 2 \sqrt{\frac{\gamma_M}{\pi}}  \int_0^\infty e^{-\delta s} \sqrt{s} \, ds  \\
&= \frac{1}{\delta}|\ln(\delta x)| + \frac{K'_1}{\delta^2} +  \frac{\sqrt{\gamma_M}}{\delta \sqrt{\delta}} < \infty,
\end{split}
\end{equation*}
where $K'_1=\max\limits_{i \in \St} |r_i-\delta+\gamma_i|$.

Therefore, $u^*=(\pi^*,c^*,I^*)$ is optimal policy of Problem \ref{prob}, and by Theorem \ref{th2}, $\hat{v}$ is the corresponding value function. $\hfill \Box$
\end{pf}

\begin{exm} \label{exm_log}
$S=2$\\
In this example, we assume there are two regimes in the economy, where regime 1 represents a bull market and regime 2 represents a bear market. According to \cite{french}, the stock returns are higher in a bull market, so $\mu_1>\mu_2$. \citet{hamilton2} found stock volatility is higher in a bear market, thus $\sigma_1<\sigma_2$. The data of overnight financing rate and treasury bill rate (see, for instance, the statistical data from Bank of Canada) suggests the risk-free interest rate is higher in good economy, hence $r_1>r_2$. \citet{haley} found the underwriting margin is negatively correlated with the interest rate, which implies the loading factor is smaller in a bull market, $\theta_1<\theta_2$. \cite{norden} observed that CDS spreads (default risk) are negatively correlated with the stock prices. Equivalently, the default risk is higher in a bear market, that is, $\eta_1<\eta_2$.

The generator matrix entries become
\[q_{11}=-\Pi_1,\,q_{12}=\Pi_1,\,q_{21}=\Pi_2,\,q_{22}=-\Pi_2,\]
with $\Pi_1,\Pi_2>0$, so the linear system \eqref{linear1} becomes
\begin{align*}
(\delta+\Pi_1)\hat{A}_1 - \Pi_1 \hat{A}_2 &= \frac{1}{\delta} (r_1+\gamma_1-\delta+\lambda_1 \hat{\Lambda}_1 )\\
- \Pi_2 \hat{A}_1 +(\delta+\Pi_2) \hat{A}_2 &= \frac{1}{\delta} (r_2+\gamma_2-\delta+\lambda_2 \hat{\Lambda}_2 )
\end{align*}
which gives a unique solution
\[ \hat{A}_i=\frac{\Pi_i(r_j+\gamma_j-\delta+\lambda_j \hat{\Lambda}_j)+(\delta + \Pi_j)(r_i+\gamma_i-\delta+\lambda_i \hat{\Lambda}_i) }{\delta^2(\delta + \Pi_1 + \Pi_2)},\]
where $i,j=1,2$ and $i\neq j$.

From the above expression of $\hat{A}_i$, we notice that only $\hat{\Lambda}_i$ is not directly given by the market. To calculate $\hat{\Lambda}_i$, we assume the loss proportion $l_t$ does not depend on time $t$ and we discuss the cases that $l$ is constant or uniformly distributed on $(0,1)$. We further assume $\frac{\theta_1}{\eta_1 (1+\theta_1)} \le \frac{\theta_2}{\eta_2 (1+\theta_2)}$. If the opposite is true, then we switch the expressions when calculating $\hat{\Lambda}_1$ and $\hat{\Lambda}_2$.

\begin{itemize}
\item[1.] $l$ is constant.\\
If $\left(\eta_i l - \frac{\theta_i}{1+\theta_i} \right)^+ \equiv 0$,
then
\[ \hat{\Lambda}_i=\ln(1-\eta_i l ).\]
Otherwise, we obtain
\[\hat{\Lambda}_i=-\ln(1+\theta_i) - \eta_i l (1+\theta_i) + \theta_i.\]

\item[2.] $l$ is uniformly distributed on $(0,1)$.\\
If $\left(\eta_i l - \frac{\theta_i}{1+\theta_i} \right)^+ \equiv 0$,
then
\[\hat{\Lambda}_i=E\left[ \ln (1-\eta_i l)\right]= \left(1-\frac{1}{\eta_i} \right) \ln(1-\eta_i)-1.\]
Otherwise, through straightforward calculus, we obtain
 \begin{align*}
 E \left[\ln \left(1-\eta_i l + \left(\eta_i l - \frac{\theta_i}{1+\theta_i}\right)^+ \right)\right] &= \left(\frac{1}{\eta_i}-1 \right)\ln(1+\theta_i) - \frac{\theta_i}{\eta_i (1+\theta_i)},\\
 E \left[ \left(\eta_i l - \frac{\theta_i}{1+\theta_i} \right)^+ \right]&=\frac{\eta_i}{2} + \frac{\theta_i^2}{2\eta_i (1+\theta_i)^2} - \frac{\theta_i}{1+\theta_i}.
 \end{align*}
 Hence,
 \[\hat{\Lambda}_i= \left(\frac{1}{\eta_i}-1 \right) \ln(1+\theta_i)- \frac{(\eta_i(1+ \theta_i) - \theta_i)^2 + 2\theta_i}{2\eta_i(1+\theta_i)}.\]
\end{itemize}
\end{exm}

\subsection{$U(y,i)=-y^\alpha, y>0, \alpha<0,\forall i \in \St$}
\label{subsec_negativepower}
In this scenario, a solution to the HJB equation \eqref{HJB2} is given by
\begin{equation}
\tilde{v}(x,i)=-\tilde{A}_i^{1-\alpha} x^\alpha,
\end{equation}
where the constants $\tilde{A}_i>0$, $i \in \St$, will be determined below.

From $\tilde{v}'(x,i)=-\alpha \tilde{A}_i^{1-\alpha} x^{-(1-\alpha)}$, $\tilde{v}''(x,i)=\alpha(1-\alpha) \tilde{A}_i^{1-\alpha} x^{-(2-\alpha)}$ and $(U')^{-1}(y,i)=(-\frac{\alpha}{y})^{\frac{1}{1-\alpha}}$, we obtain
\begin{equation*}
\pi^*(x,i)=\frac{\mu_i-r_i}{(1-\alpha)\sigma_i^2}\quad \text{and} \quad c^*(x,i)=\frac{x}{\tilde{A}_i}>0.
\end{equation*}

Solving $(1+\theta_i)\tilde{v}'(x,i)=\tilde{v}'(x-d_i,i)$ gives $d_i=\nu_i x$, where $\nu_i:= 1-(1+\theta_i)^{-\frac{1}{1-\alpha}}$. Then
\[ I^*(x,i,l)=(\eta_i l - \nu_i)^+ x.\]

By plugging the candidate policy into the HJB equation \eqref{HJB2}, we find the constants $\tilde{A}_i$ should satisfy the following non-linear system
\begin{equation} \label{nonlinear}
\left(\delta  - \alpha r_i - \frac{\alpha}{1-\alpha} \gamma_i  + \lambda_i(1-\tilde{\Lambda}_i) \right) \tilde{A}_i^{1-\alpha} - (1-\alpha) \tilde{A}_i^{-\alpha} = \sum_{j \in \St} q_{ij} \tilde{A}_j^{1-\alpha},
\end{equation}
where $\tilde{\Lambda}_i:=E \big[(1-\eta_i l +(\eta_i l - \nu_i)^+)^\alpha \big] - \alpha (1+\theta_i) E \big[(\eta_i l - \nu_i)^+\big]$.

In order to guarantee the above non-linear system has a unique positive solution, we need the following technical condition
\begin{equation}
\delta > \max_{i\in \St} \left\{\alpha r_i + \frac{\alpha}{1-\alpha} \gamma_i  - \lambda_i(1-\tilde{\Lambda}_i) \right\}. \label{tech}
\end{equation}

\begin{lem}
The non-linear system \eqref{nonlinear} has a unique positive solution $\tilde{A}_i$, $i \in \St$, if the condition \eqref{tech} holds.
\end{lem}

\begin{pf}
See Lemma 4.1 in \citet{sotomayor}. \hfill$\Box$
\end{pf}

\begin{prop}
The function $\tilde{v}=\tilde{v}(\cdot,\cdot)$, given by
\[ \tilde{v}(x,i)=\begin{cases}
-\tilde{A}_i^{1-\alpha}x^\alpha, & x>0\\
-\infty, &x=0 \end{cases}, \]
where $\tilde{A}_i$ is the unique solution to the non-linear system \eqref{nonlinear}, is the value function of Problem \ref{prob}. Furthermore, the policy given by
\[ u^*(t)=\left( \frac{\mu_{\epsilon(t)}-r_{\epsilon(t)}}{(1-\alpha)\sigma_{\epsilon(t)}^2}, \frac{X_t^*}{\tilde{A}_{\epsilon(t)}}, \left(\eta_{\epsilon(t)} \, l_t - \nu_{\epsilon(t)} \right)^+X_t^* \right)\]
is optimal policy of Problem \ref{prob}.
\end{prop}

\begin{pf}
To verify that the candidate policy is admissible, we consider
an upper bound process $\tilde{Z}$ of $X^*$ with the dynamics
\[\frac{d\tilde{Z}_t}{\tilde{Z}_t}= \left( r_{\epsilon(t)} + \frac{(\mu_{\epsilon(t)}-r_{\epsilon(t)})^2}{(1-\alpha) \sigma_{\epsilon(t)}^2} \right)dt + \frac{\mu_{\epsilon(t)}-r_{\epsilon(t)}}{(1-\alpha)\sigma_{\epsilon(t)}} dW_t. \]

Given $\tilde{Z}_0=X^*_0=x$, we can solve the above SDE to obtain
\[ \tilde{Z}_t = x \exp \left\{ \int_0^t \Big( r_{\epsilon(s)} + \frac{(1-2\alpha)(\mu_{\epsilon(s)}-r_{\epsilon(s)})^2}{2(1-\alpha)^2 \sigma_{\epsilon(s)}^2} \Big) ds + \int_0^t \frac{\mu_{\epsilon(s)}-r_{\epsilon(s)}}{(1-\alpha)\sigma_{\epsilon(s)}} dW_s \right\}.\]

We use this upper bound process $\tilde{Z}$ to verify that the conditions for an admissible control are satisfied.  We have for every $t\ge 0$ that
\begin{align*}
E_{x,i} \left[ \int_0^t \sigma_{\epsilon(t)}^2 (\pi^*_s)^2 ds\right] &\le \frac{2\gamma_M \, t}{(1-\alpha)^2 } < \infty, \\
E_{x,i} \left[ \int_0^t c^*_s ds\right]&\le \frac{1}{\tilde{A}_m} E_{x,i} \left[ \int_0^t \tilde{Z}_s ds \right] \\
&\le \frac{x}{\tilde{A}_m} \int_0^t e^{K_2 s} ds <\infty,\\
E_{x,i} \left[ \int_0^\Theta e^{-\delta t} U^+(c_t^*) dt \right] &\le E_{x,i} \left[ \int_0^\Theta e^{-\delta t} \left(-(\frac{\tilde{Z}_t}{\tilde{A}_{\epsilon(t)}})^\alpha \right)^+ dt \right]=0,
\end{align*}
where $\tilde{A}_m:=\min\limits_{i \in \St}\{\tilde{A}_i\}$ and $K_2=\max\limits_{i\in \St} \left\{r_i+\frac{2 \gamma_i}{1-\alpha}\right\}$.

Besides, we can verify that $I^*_t \in \mathcal{I}_t$ since $0\le I^*_t(Y)=(Y - d_{\epsilon(t)})^+ \le Y$, for every $\F_t$-measurable random variable $Y$.

Therefore, $u^*$ defined above is admissible and then is optimal policy of Problem \ref{prob}. By definition, smooth function $\tilde{v}(\cdot,i)$ is strictly increasing and concave, and satisfies $\tilde{v}(0,i)=-\infty$, $\forall i \in \St$. From the construction of $\tilde{A}_i$, the HJB equation \eqref{HJB2} holds for all $i \in \St$. Therefore, according to Theorem \ref{th2}, $\tilde{v}$ is the value function of Problem \ref{prob}.  \hfill $\Box$
\end{pf}

\begin{exm} \label{exm_negativepower}
$S=2$\\
To solve the non-linear system \eqref{nonlinear}, we need to find $\tilde{\Lambda}_i$ first. In this example, we show how to find $\tilde{\Lambda}_i$ when $l$ is constant or is uniformly distributed on $(0,1)$. Without loss of generality, we assume $\frac{\nu_1}{\eta_1 } \le \frac{\nu_2}{\eta_2 }$. If the opposite holds, we switch the formulas for $\tilde{\Lambda}_1$ and $\tilde{\Lambda}_2$. The results will be used for economic analysis in the next section.
\begin{itemize}
\item[1.] $l$ is constant.\\
If $(\eta_i l - \nu_i)^+ \equiv 0$,
then
\[\tilde{\Lambda}_i=(1-\eta_i l)^\alpha.\]
Otherwise, we obtain
\[\tilde{\Lambda}_i=(1-\nu_i)^\alpha - \alpha(1+\theta_i)(\eta_i l - \nu_i).\]

\item[2.] $l$ is uniformly distributed on $(0,1)$.\\
If $(\eta_i l - \nu_i)^+ \equiv 0$,
then
\[\tilde{\Lambda}_i=E\left[(1-\eta_i l)^\alpha\right]=
\begin{cases}
-\frac{1}{\eta_i} \ln(1-\eta_i), &\alpha=-1\\
\frac{1}{\eta_i(1+\alpha)} (1-(1-\eta_i)^{1+\alpha}), &\alpha \neq -1
\end{cases}. \]
Otherwise, we obtain
\begin{equation*}
 E[(\eta_i l - \nu_i)^+]=\int_{\frac{\nu_i}{\eta_i}}^1 (\eta_i l -\nu_i) dl =\frac{(\eta_i-\nu_i)^2}{2\eta_i},
\end{equation*}
and when $\alpha=-1$,
\[E[(1-\eta_i l + (\eta_i l - \nu_i)^+)^\alpha] =(1-\nu_i)^{-1} \left(1-\frac{\nu_i}{\eta_i} \right) - \frac{1}{\eta_i} \ln(1-\nu_i),\]
and when $\alpha \neq -1$,
\[E[(1-\eta_i l + (\eta_i l - \nu_i)^+)^\alpha] = (1-\nu_i)^\alpha \left(1-\frac{\nu_i}{\eta_i}-\frac{1-\nu_i}{\eta_i(1+\alpha)}\right) + \frac{1}{\eta_i(1+\alpha)}.\]
Therefore, if $(\eta_i l - \nu_i)^+ \not\equiv 0$,
and $\alpha=-1$, then
\[\tilde{\Lambda}_i=(1-\nu_i)^{-1} \left(1-\frac{\nu_i}{\eta_i}\right) - \frac{1}{\eta_i} \ln(1-\nu_i)+(1+\theta_i)\frac{(\eta_i-\nu_i)^2}{2\eta_i};\]
and if $(\eta_i l - \nu_i)^+ \not\equiv 0$,
and $\alpha \neq -1$, then
\[\tilde{\Lambda}_i=(1-\nu_i)^\alpha \left(1-\frac{\nu_i}{\eta_i}-\frac{1-\nu_i}{\eta_i(1+\alpha)}\right) + \frac{1}{\eta_i(1+\alpha)}-\alpha(1+\theta_i)\frac{(\eta_i-\nu_i)^2}{2\eta_i}.\]

\end{itemize}
\end{exm}

\subsection{$U(y,i)=y^\alpha, y>0, 0<\alpha<1, \forall \, i \in \St$}
\label{subsec_positivepower}

In this case, a solution to the HJB equation \eqref{HJB1} has the form
\begin{equation}
\bar{v}(x,i)=\bar{A}_i^{1-\alpha}x^\alpha,
\end{equation}
where the constants $\bar{A}_i>0$, $i \in \St$, will be determined below.

Then we can find the candidate for $\pi^*$ and $c^*$ as
\begin{equation*}
\pi^*(x,i)=\frac{\mu_i-r_i}{(1-\alpha) \sigma_i^2} \quad \text{and} \quad c^*(x,i)=\frac{x}{\bar{A}_i}.
\end{equation*}

From $(1+\theta_i)\bar{v}'(x,i)=\bar{v}'(x-d_i,i)$, we can solve to obtain $d_i=\nu_i x$ with $\nu_i:= 1-(1+\theta_i)^{-\frac{1}{1-\alpha}}$. By Theorem \ref{opin}, we have
\[ I^*(x,i)=(\eta_i l - \nu_i)^+ x.\]

Plugging the candidate policy into the HJB equation \eqref{HJB1} yields
\begin{equation} \label{nonlinear1}
\left(\delta - \alpha r_i - \frac{\alpha}{1-\alpha} \gamma_i + \lambda_i (1 - \bar{\Lambda}_i) \right) \bar{A}_i^{1-\alpha} - (1-\alpha) \bar{A}_i^{-\alpha} = \sum_{j \in \St} q_{ij} \bar{A}_j^{1-\alpha},
\end{equation}
where $\bar{\Lambda}_i:=E \big[(1-\eta_i l +(\eta_i l - \nu_i)^+)^\alpha\big] - \alpha (1+\theta_i) E \big[(\eta_i l - \nu_i)^+ \big]$.

We need to impose an extra requirement for $\delta$
\begin{equation} \label{tech2}
\delta > \max_{i \in \St} \left\{\alpha r_i + \frac{\alpha}{1-\alpha} \gamma_i  \right\}.
\end{equation}

\begin{lem}
The non-linear system \eqref{nonlinear1} has a unique positive solution $\bar{A}_i$, $i \in \St$, if the condition \eqref{tech2} is satisfied.
\end{lem}

\begin{pf}
See Lemma 4.2 in \citet{sotomayor}. \hfill$\Box$
\end{pf}

\begin{prop}
The function $\bar{v}(x,i)=\bar{A}_i^{1-\alpha} x^\alpha$, $x\ge 0$, where $\bar{A}_i$ is the unique solution to the non-linear system \eqref{nonlinear1}, is the value function of Problem \ref{prob}. Furthermore, the policy given by
\[ u^*(t):=\left( \frac{\mu_{\epsilon(t)}-r_{\epsilon(t)}}{(1-\alpha)\sigma_{\epsilon(t)}^2}, \frac{X^*_t}{\bar{A}_{\epsilon(t)}},(\eta_{\epsilon(t)} \, l_t - \nu_{\epsilon(t)})^+ X^*_t\right) \]
is optimal policy of Problem \ref{prob}.
\end{prop}

\begin{pf}
We use the same upper bound process $\tilde{Z}$ defined in Section \ref{subsec_negativepower}. By following a similar argument as in the previous proposition, we can easily verify $E_{x,i} \left[ \int_0^t \sigma_{\epsilon(t)}^2 (\pi^*_s)^2 ds \right] < \infty$, $E_{x,i} \left[ \int_0^t c_s^* ds \right] < \infty$ and $I^*_t \in \mathcal{I}_t$, $\forall t \ge 0$.

Besides, we have
\begin{align*}
E_{x,i} \left[\int_0^\Theta e^{-\delta t}U^+(c^*_t) dt \right] &\le E_{x,i} \left[\int_0^\infty e^{-\delta t} \left( \frac{\tilde{Z}_t}{\bar{A}_{\epsilon(t)}} \right)^\alpha dt \right]\\
&= \frac{x^\alpha}{\bar{A}_m^\alpha} \int_0^\infty  e^{-\delta t} \exp \left(\int_0^t \Big(\alpha r_{\epsilon(s)} + \frac{\alpha}{1-\alpha} \gamma_{\epsilon(s)} \Big)ds \right)dt \\
&\le\frac{x^\alpha}{K_3 \bar{A}_m^\alpha}<\infty,
\end{align*}
where $\bar{A}_m=\min\limits_{i \in \St} \bar{A}_i$ and $K_3=\min\limits_{i \in \St} (\delta-\alpha r_i-\frac{\alpha}{1-\alpha}\gamma_i)>0$ ($K_3>0$ is because of the condition \eqref{tech2}).

By definition, $\bar{v}(\cdot,i)\in C^2(0,\infty)$ is strictly increasing and concave, and satisfies $\bar{v}(0,i)=\frac{U(0,i)}{\delta}=0$ for all $i\in \St$. By the construction of constants $\bar{A}_i$, the HJB equation \eqref{HJB2} holds for all $i \in \St$.

Therefore, $u^*$ is admissible and then is optimal policy of Problem \ref{prob}. Furthermore, by Theorem \ref{th1}, $\bar{v}$ defined above is the value function of Problem \ref{prob}. \hfill $\Box$
\end{pf}

\begin{exm}
\label{exm_positivepower}
$S=2$\\
We notice that the non-linear systems \eqref{nonlinear} and \eqref{nonlinear1} are identical expect that $\alpha$ is negative in \eqref{nonlinear} while in \eqref{nonlinear1}, $\alpha \in (0,1)$. Hence in a two-regime economy, we shall obtain $\bar{\Lambda}_i$ in the same form of $\tilde{\Lambda}_i$ as in Example \ref{exm_negativepower}.
\end{exm}

\subsection{$U(y,i)=\beta_i y ^{1/2}, y>0, \beta_i>0,i=1,2$}
In this case, a solution to the HJB equation \eqref{HJB1} is given by
\begin{equation}
\check{v}(x,i)=(\check{A}_i x)^{1/2},
\end{equation}
where the constants $\check{A}_i>0, i=1,2$, will be determined below.

From $\check{v}'(x,i)=\frac{1}{2} \check{A}_i^{\frac{1}{2}} x^{-\frac{1}{2}}$, $\check{v}''(x,i)=-\frac{1}{4} \check{A}_i^{\frac{1}{2}} x^{-\frac{3}{2}}$ and $(U')^{-1}(y,i)=(\frac{\beta_i}{2y})^2$, we obtain the candidate for $\pi^*$ and $c^*$
\[\pi^*(x,i)=\frac{2(\mu_i-r_i)}{\sigma_i^2} \text{ and } c^*(x,i)=\frac{\beta_i^2 x}{\check{A}_i}.\]

Solving $(1+\theta_i) \check{v}'(x,i)=\check{v}'(x-d_i,i)$ gives $d_i=\check{\nu}_i x$ where $\check{\nu}_i:=1-\frac{1}{(1+\theta_i)^2}$. Thus a candidate for optimal insurance is
\[I^*(x,i)=(\eta_i l - \check{\nu}_i)^+x.\]

From the HJB equation \eqref{HJB1}, we obtain the following nonlinear system
\begin{equation*}
\left(\delta - \frac{1}{2}r_i - \gamma_i + \lambda_i(1-\check{\Lambda}_i) \right) \check{A}_i^{1/2} -\frac{1}{2} \frac{\beta_i^2}{\check{A}_i^{1/2}}=\sum_{j \in \St} q_{ij} \check{A}_j^{1/2},
\end{equation*}
where $\check{\Lambda}_i:=E\left[(1-\eta_i l + (\eta_i l - \check{\nu}_i)^+)^{1/2}\right] - \frac{1}{2}(1+\theta_i)E\left[(\eta_i l - \check{\nu}_i)^+\right]$.

Since $S=2$, so we have $q_{11}=-\Pi_1,\,q_{12}=\Pi_1,\,q_{21}=\Pi_2,\,q_{22}=-\Pi_2$ with $\Pi_1,\Pi_2>0$. Thus we can rewrite the above system as
\begin{equation}\label{nonlinear2}
\check{\xi}_i \check{A}_i - \frac{\beta_i^2}{2\Pi_i}=(\check{A}_1 \check{A}_2)^{1/2},
\end{equation}
where $\check{\xi}_i:=\frac{1}{\Pi_i} \left[\delta + \Pi_i - \frac{1}{2}r_i-\gamma_i + \lambda_i(1-\check{\Lambda}_i) \right]$.

\begin{lem}
The non-linear system \eqref{nonlinear2} has a real solution $\check{A}_i \ge \dfrac{\beta_i^2}{2\Pi_i \check{\xi}_i} >0$, $i=1,2$, if $\delta > \max\limits_{i=1, 2} \left\{\frac{1}{2} r_i + \gamma_i, \frac{1}{2} r_i + \gamma_i  - \lambda_i(1-\check{\Lambda}_i) \right\}$.
\end{lem}

\begin{pf}
The non-linear system \eqref{nonlinear2} is equivalent to
\begin{equation*}
\check{\xi}_1 \check{A}_1 - \frac{\beta_1^2}{2\Pi_1} = \sqrt{\check{A}_1 \check{A}_2} = \check{\xi}_2 \check{A}_2 - \frac{\beta_2^2}{2\Pi_2}.
\end{equation*}

Solving this system for $\check{A}_1$ gives
\[ \left( \frac{\check{\xi}_1}{\check{\xi}_2} - \check{\xi}_1^2\right) \check{A}_1^2 - \left( \frac{\beta_1^2}{2\Pi_1 \check{\xi}_2} - \frac{\beta_2^2}{2\Pi_2 \check{\xi}_2} - \frac{\check{\xi}_1 \beta_1^2}{\Pi_1} \right) \check{A}_1 - \frac{\beta_1^4}{4\Pi_1^2}=0. \]
The discriminant of the above quadratic equation is
\[ \Delta=\left(\frac{\beta_1^2}{2\Pi_1 \check{\xi}_2}-\frac{\beta_2^2}{2\Pi_2 \check{\xi}_2} \right)^2 + \frac{\check{\xi}_1 \beta_1^2 \beta_2^2}{\Pi_1 \Pi_2 \check{\xi}_2}.\]

Since $\delta > \frac{1}{2} r_i + \gamma_i  - \lambda_i(1-\check{\Lambda}_i)$, we have $\check{\xi}_i>1$, $i=1,2$ and then $\Delta > 0$, which implies $\check{A}_1$ has a real solution. Besides, $\sqrt{\check{A}_1 \check{A}_2} \ge 0$, so $\check{A}_1 \ge \dfrac{\beta_1^2}{2\Pi_1 \check{\xi}_1} >0$. Similar analysis also applies to $\check{A}_2$. $\hfill \Box$
\end{pf}

\begin{prop}
The function $\check{v}$ defined by $\check{v}(x,i)=(\check{A}_i x)^{1/2}$, $x > 0$, where $\check{A}_i$ is the positive solution to the non-linear system \eqref{nonlinear2}, is the value function of Problem \ref{prob}. Furthermore, the policy given by
\[ u^*(t):= \left( \frac{2(\mu_{\epsilon(t)} - r_{\epsilon(t)})}{\sigma_{\epsilon(t)}^2}, \frac{\beta_{\epsilon(t)}^2 X_t^*}{\check{A}_{\epsilon(t)}}, (\eta_{\epsilon(t)} \, l_t - \check{\nu}_{\epsilon(t)})^+ X_t^* \right)\]
is optimal policy of Problem \ref{prob}.
\end{prop}

\begin{pf}
We consider an upper bound process $\check{Z}$ of $X^*$ to verify that the candidate policy is admissible. The dynamics of $\check{Z}$ is given by
\[ \frac{d \check{Z}_t} {\check{Z}_t} = \left( r_{\epsilon(t)} + \frac{2(\mu_{\epsilon(t)}-r_{\epsilon(t)})^2}{\sigma_{\epsilon(t)}^2}\right) dt + \frac{2(\mu_{\epsilon(t)} - r_{\epsilon(t)})}{\sigma_{\epsilon(t)}} dW(t), \]
with initial condition $\check{Z}_0=X^*_0=x$.

The solution to the above SDE is
\[ \check{Z}_t=x \cdot \exp\left\{ \int_0^t r_{\epsilon(s)} ds + 2\int_0^t \frac{\mu_{\epsilon(s)} - r_{\epsilon(s)}}{\sigma_{\epsilon(s)}} dW(s)\right\}.\]

Since $X^*_t \le \check{Z}_t$, $\forall \, t\ge 0$, we have
\begin{align*}
E_{x,i} \left[ \int_0^t c^*(s) ds\right] &\le \frac{\beta_M^2}{\check{A}_m} E_{x,i} \left[ \int_0^t \check{Z}_s ds \right]\\
&= \frac{\beta_M^2 x}{\check{A}_m} \int_0^t \exp\left( \int_0^s (r_{\epsilon(v)}+ 4 \gamma_{\epsilon(v)}) dv\right) ds\\
&\le  \frac{\beta_M^2 x}{\check{A}_m}  \int_0^t e^{K_4 s} = \frac{\beta_M^2 x}{K_4 \check{A}_m} (e^{K_4 t}-1)<\infty,
\end{align*}
where $\beta_M=\max\{ \beta_1, \beta_2\}$, $\check{A}_m=\min\{ \check{A}_1,\check{A}_2\}$ and $K_4=\max\limits_{i=1,2}\{r_i + 4\gamma_i\}$.

Furthermore, we calculate
\begin{align*}
E_{x,i} \left[ \int_0^\Theta e^{-\delta s} U^+(c^*_s,\epsilon_s) ds \right] &\le E_{x,i} \left[ \int_0^\infty e^{-\delta s} \beta_{\epsilon(s)} \frac{\beta_{\epsilon(s)}(X^*_s)^{1/2}}{\check{A}_{\epsilon(s)}^{1/2}} ds \right]\\
&\le E_{x,i} \left[ \int_0^\infty e^{-\delta s} \beta_{\epsilon(s)} \frac{\beta_{\epsilon(s)} \check{Z}_s^{1/2}}{\check{A}_{\epsilon(s)}^{1/2}} ds \right]\\
&= \frac{\beta_M^2 x^{1/2}}{\check{A}_m^{1/2}} \int_0^t e^{-\delta s} \exp \left(\int_0^s (\frac{1}{2} r_{\epsilon(v)} + \gamma_{\epsilon(v)})dv  \right) ds\\
&\le \frac{\beta_M^2 x^{1/2}}{\check{A}_m^{1/2}} \int_0^\infty e^{-K_5 s}ds = \frac{\beta_M^2 x^{1/2}}{K_5 \check{A}_m^{1/2}}  < \infty,
\end{align*}
where $K_5:=\min_{i \in \St} (\delta - \frac{1}{2}r_i - \gamma_i)>0$ (Notice $K_5>0$ due to the assumption that $\delta >\frac{1}{2} r_i + \gamma_i$, $\forall \, i \in \St$).

Besides, $\forall \, t\ge 0$,
\[E_{x,i} \left[ \int_0^t \sigma_{\epsilon(s)}^2 (\pi^*_s)^2 ds \right] \le 8 \gamma_M t< +\infty, \]
and $0\le I^*_t(Y) = (Y -\check{\nu}_{\epsilon(t)} X_t^*)^+ \le Y$, for every $\F_t$-measurable random variable $Y$.

We have proved $u^*$ is admissible and thus $u^*$ is optimal policy of Problem \ref{prob}. By definition, $\check{v}(\cdot,i) \in C^2(0,\infty)$ is strictly increasing and concave, and satisfies $\check{v}(0,i)=\frac{U(0,i)}{\delta}$, $i=1,2$. By the construction of $\check{A}_i$, the HJB equation \eqref{HJB1} is satisfied for $i=1,2$.
Therefore, by Theorem \ref{th1}, the value function is given by $\check{v}(x,i)+\frac{1}{\delta} E_{x,i} [ \int_0^\infty e^{-\delta s} dU(0,\epsilon_s) ] = \check{v}(x,i)$ because $dU(0,\epsilon_s)=0$. $\hfill \Box$
\end{pf}

\section{Economic Analysis}
\label{sec_economicanalysis}
In this section, we analyze the impact of market parameters and the investor's risk aversion on optimal policy, and how insurance affects the expected total discounted utility of consumption (the value function). To conduct the economic analysis, we assume there are two regimes in the economy, like in Example \ref{exm_log}, Example \ref{exm_negativepower}, and Example \ref{exm_positivepower}: regime 1 represents a bull market while regime 2 represents a bear market. We only consider the first three utility functions in the economic analysis.

\subsection{Impact of market parameters and risk aversion on optimal policy}
\label{impact1}
According to the results obtained in Section \ref{sec_explicitsolution}, we write the optimal proportion invested in the stock in an uniform expression
\begin{equation} \label{opp}
\pi^*_t=\frac{1}{1-\alpha} \frac{\mu_{\epsilon(t)}-r_{\epsilon(t)}}{\sigma_{\epsilon(t)}^2},
\end{equation}
where $\alpha=0$ when $U(y,i)=\ln(y)$.

During any given regime, the optimal investment proportion in the stock $\pi^*$ is constant, and only depends on market parameters (expected excess return over variance) and the investor's risk aversion parameter $\alpha$.

The dependency of $\pi^*$ on market parameters is evident. Through empirical research, \citet{french} find that the expected excess return over variance is higher in good economy. Therefore, in a bull market, investors should invest a greater proportion of their wealth on the stock.

Expression \eqref{opp} shows that $\pi^*$ is inversely proportional to the relative risk aversion $1-\alpha$, so low risk-averse investors (with greater $\alpha$) will invest a higher proportion of their wealth on the stock.

For all three cases, the optimal consumption rate process is proportional to the wealth process and such ratio $\kappa(t):=\frac{c^*(t)}{X^*(t)}$ is given by
\begin{equation*}
\kappa(t)=
\begin{cases}
\delta, &\text{ if } U(y,i)=\ln(y),\alpha=0;\\
\dfrac{1}{\tilde{A}_{\epsilon(t)}}, &\text{ if } U(y,i)=-y^\alpha,\alpha<0;\\
\dfrac{1}{\bar{A}_{\epsilon(t)}}, &\text{ if } U(y,i)=y^\alpha,0<\alpha<1.
\end{cases}
\end{equation*}
Since $\kappa(t)$ is positive in all three cases, investors will consume proportionally more when they become wealthier. To examine the dependency of the optimal consumption to wealth ratio $\kappa(t)$ on $\alpha$, we separate our discussion into the following three cases.

For moderate risk-averse investors ($\alpha=0$), $\kappa(t)$ is constant regardless of the market regimes, so moderate risk-averse investors consume the same proportion of their wealth in both bull and bear markets.

\begin{figure}[ht!]
 \centering
 \includegraphics[width=130mm]{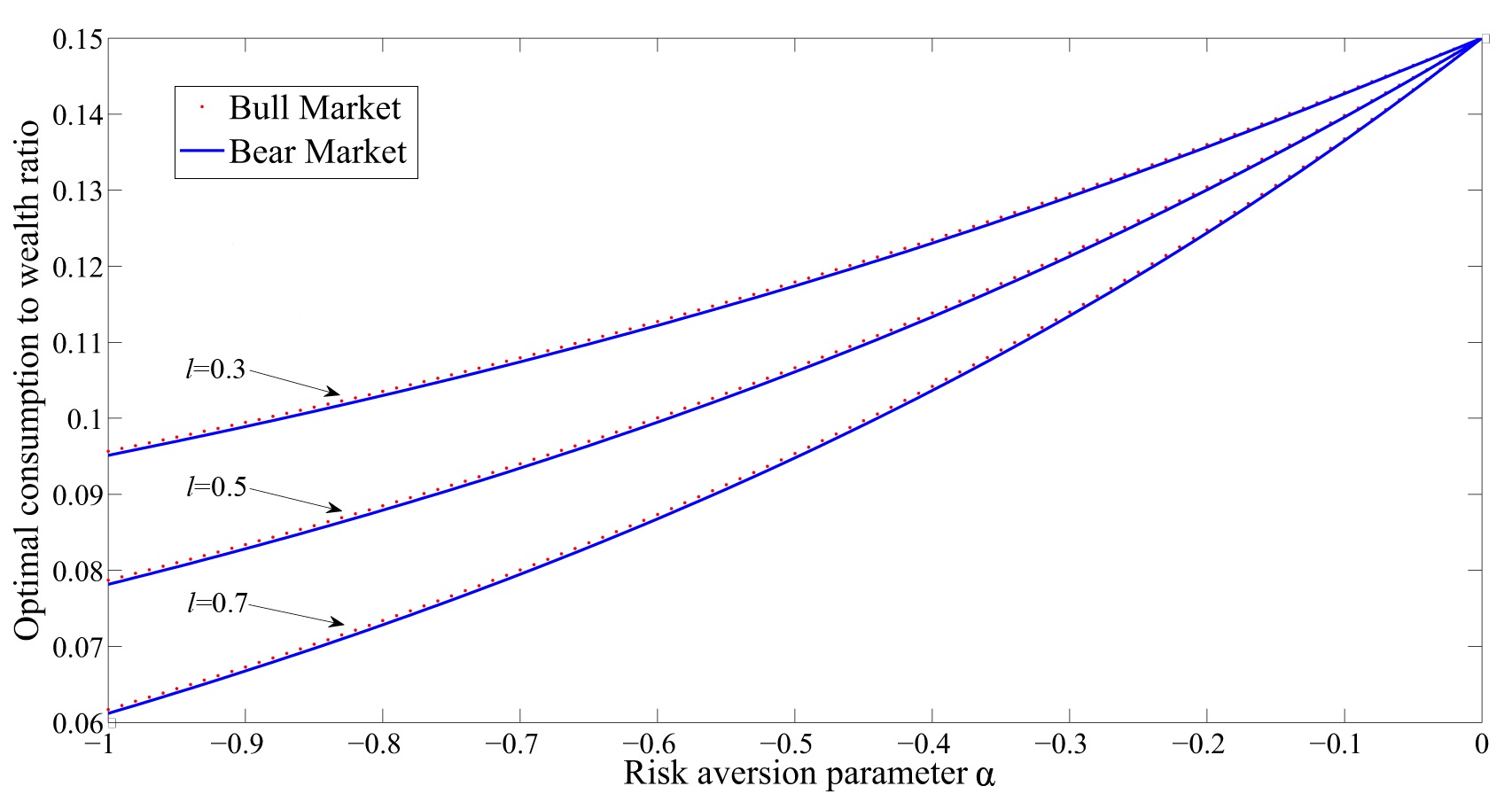}
 \caption{Optimal consumption to wealth ratio when $\alpha<0$}
 \label{pic_consumption_a<0}
\end{figure}

For high risk-averse investors ($\alpha <0$), their optimal consumption to wealth ratio is given by $1/\tilde{A}_i,i=1,2$, where $\tilde{A}$ can be obtained from the system \eqref{nonlinear}. To find a numerical solution to the system \eqref{nonlinear}, we set market parameters as $\mu_1=0.2, \mu_2=0.15, r_1=0.08, r_2=0.03, \sigma_1=0.25, \sigma_2=0.6,  \theta_1=0.15, \theta_2=0.25, \eta_1=0.8, \eta_2=1, \lambda_1=0.1, \lambda_2=0.2, \Pi_1=6.04, \Pi_2=6.4$, and $\delta=0.15$ (for the convenience of citation thereafter, we denote the choice of market parameters here as Parameter Set I). Notice that these parameters satisfy the technical condition \eqref{tech}. We draw graphs in Figure \ref{pic_consumption_a<0} for the optimal consumption to wealth ratio when $-1<\alpha<0$ and $l=0.3$, $l=0.5$, and $l=0.7$. We see that the optimal consumption to wealth ratio is an increasing function of $\alpha$.
Thus, the higher the risk tolerance, the higher the proportion of consumption over wealth.
For the above parameter values, we find $1/\tilde{A}_1 > 1/\tilde{A}_2$, which can be seen from Figure \ref{pic_consumption_a<0}.
Hence investors should allocate a higher proportion of their wealth to consumption in a bull market. For any chosen investor (fixed $\alpha$), she/he will behave more conservatively by reducing the proportion spent in consumption when facing larger losses (greater $l$). This behavior was not noticed in \citet{sotomayor}, because they did not incorporate an insurable loss in their model.
Besides, from a mathematical point of view, the ratios all converge to $0.15$ when $\alpha$ approaches $0$, which is exactly the same optimal consumption to wealth ratio when $\alpha=0$ ($\delta=0.15$).

For low risk-averse investors ($0<\alpha<1$), the optimal consumption to wealth ratio is given by $1/\bar{A}_i,i=1,2,$ where $1/\bar{A}_i$ can be calculated from the system \eqref{nonlinear1}.
We set market parameters to be $\mu_1=0.2, \mu_2=0.15, r_1=0.15, r_2=0.1, \sigma_1=0.4, \sigma_2=0.6,  \theta_1=0.15, \theta_2=0.25, \eta_1=0.8, \eta_2=1, \lambda_1=0.1, \lambda_2=0.2, \Pi_1=6.04, \Pi_2=6.4$, and $\delta=0.2$ (denoted as Parameter Set II). For these parameters values, the corresponding technical condition \eqref{tech2} is satisfied.
\begin{figure}[htb!]
 \centering
 \includegraphics[width=130mm]{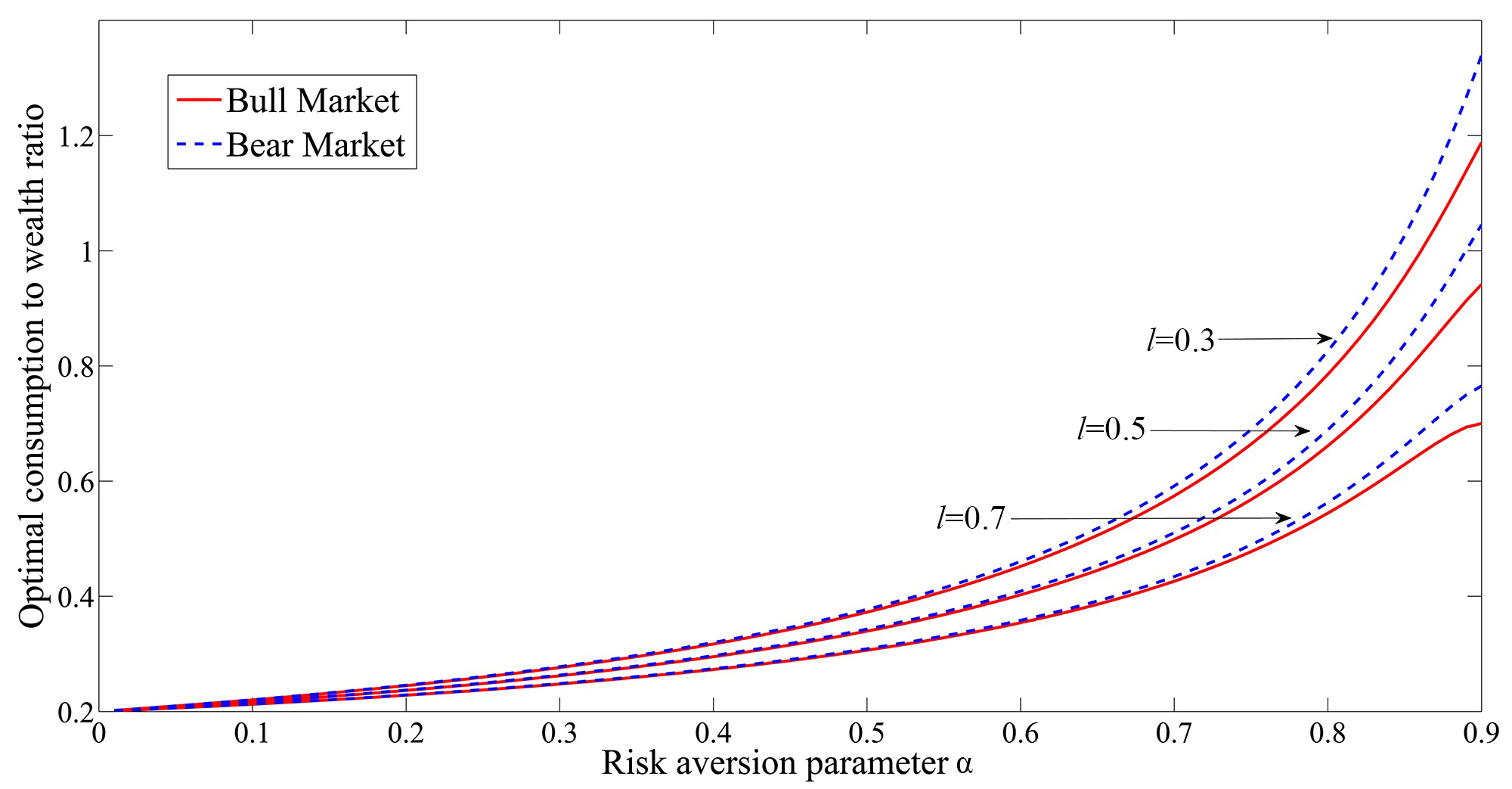}
 \caption{Consumption to wealth ratio when $0<\alpha<1$}
 \label{pic_consumption_a>0}
\end{figure}
Figure \ref{pic_consumption_a>0} shows that the optimal consumption to wealth ratio when $l=0.3$, $0.5$, and $0.7$. Similar to the previous case, we also observe that the optimal consumption to wealth ratio is an increasing function of $\alpha$. However, contrary to the previous case, we have $1/\bar{A}_1<1/\bar{A}_2$ when $0<\alpha<1$. This means low risk-averse
investors ($0<\alpha<1$) spend a smaller proportion of their wealth on consumption in a bull market.
We notice that for very low risk-averse investors ($\alpha$ close to 1), the optimal consumption to wealth ratio is even greater than 1, meaning they finance consumption by borrowing.

By comparing all three cases, we conclude that investors with high risk tolerance ($\alpha$ is large) consume at large proportion of their wealth in every market regime.
However, investors' consumption decision depends on the market regimes, and investors with different risk aversion attitudes behave differently in bull and bear markets.

The optimal insurance for all three utility functions is deductible insurance, and is given by
\begin{equation*}
I^*_t=
\left[ \eta_{\epsilon(t)} \, l  -  1+(1+\theta_{\epsilon(t)})^{-\frac{1}{1-\alpha}}  \right]^+ X_t^*.
\end{equation*}
We observe that, for each fixed regime, the optimal insurance is proportional to the investor's wealth $X^*$.
We note that it is optimal to buy insurance if and only if
$\eta_{\epsilon(t)} \, l  -  1 + (1+\theta_{\epsilon(t)})^{-\frac{1}{1-\alpha}} >0$,
or equivalently if and only if
\begin{equation}
\eta_{\epsilon(t)} l  >  1-(1+\theta_{\epsilon(t)})^{-\frac{1}{1-\alpha}}.
\label{condition-to-buy-insurance}
\end{equation}
Thus, it is optimal to buy insurance if and only if,
relative to the other variables,
$\eta_{\epsilon(t)}$ is large,
$l$ is large,
$\theta_{\epsilon(t)}$ is small,
and
$\alpha$ is small (we recall that $\alpha \in (-\infty,1)$).
That is, it is optimal to buy insurance if the insurable loss is large, the cost of insurance is low, and the investor is very risk averse.
It is surprising that the variable
$\lambda_{\epsilon(t)}$ does not appear explicitly in this expression.
Our explanation is that  $\lambda_{\epsilon(t)}$ is implicitly incorporated in
$X_t^*$, so $\lambda_{\epsilon(t)}$   is important as well
 to determine the optimal insurance.

If it is optimal to buy insurance, or equivalently, the condition \eqref{condition-to-buy-insurance} is satisfied,
then $I^*_t=
\left[ \eta_{\epsilon(t)} \, l  -  1+(1+\theta_{\epsilon(t)})^{-\frac{1}{1-\alpha}}  \right]X_t^*.$
Thus, as expected, the optimal insurance is proportional to
$\eta_{\epsilon(t)}$ and
$l$.
Furthermore,
\begin{align*}
\frac{\partial I_t^*}{\partial \theta_{\epsilon(t)}}
&=-\left[\frac{1}{1-\alpha}(1+\theta_{\epsilon(t)})^{-\frac{2-\alpha}{1-\alpha}}\right]X_t^* <0, \\
\frac{\partial^{2} I_t^*}{\partial \theta_{\epsilon(t)}^{2}}&=\left[ \frac{2-\alpha}{(1-\alpha)^2}
(1+\theta_{\epsilon(t)})^{\frac{2\alpha -3}{1-\alpha}}\right]X_t^* >0.
\end{align*}
Hence, the optimal insurance is a decreasing and convex function of
$\theta$.
The decreasing property means that, as the premium loading $\theta$ increases, it is optimal to reduce
the purchase of insurance. The convexity indicates the amount of reduction in insurance decreases as
the premium loading rises.

In addition, if it is optimal to buy insurance (when condition \eqref{condition-to-buy-insurance} is satisfied),
then
\begin{align*}
\frac{\partial I_t^*}{\partial \alpha}&=-\left[
\frac{1}{(1-\alpha)^2} \ln(1+\theta_{\epsilon(t)}) (1+\theta_{\epsilon(t)})^{-\frac{1}{1-\alpha}}  \right] X_t^*<0,\\
\frac{\partial^{2} I_t^*}{\partial \alpha^{2}}&=\left[\frac{\ln(1+\theta_{\epsilon(t)})}{(1-\alpha)^3} \left( \frac{\ln(1+\theta_{\epsilon(t)})}{1-\alpha}-2 \right) (1+\theta_{\epsilon(t)})^{-\frac{1}{1-\alpha}}\right]  X_t^* .
\end{align*}
Hence, the optimal insurance is a decreasing function of
$\alpha$, which implies the higher the risk tolerance, the smaller amount spent on insurance. We observe $\frac{\partial^{2} I_t^*}{\partial \alpha^{2}}$ and $ \frac{\ln(1+\theta_{\epsilon(t)})}{1-\alpha}-2$ have the same sign. Recall that $\theta$ is the premium loading, which usually does not exceed $100\%$. So when $\alpha \le 0$, we have $\frac{\partial^{2} I_t^*}{\partial \alpha^{2}}<0$. This indicates that for high and moderate risk-averse investors ($\alpha \le 0$), the reduction in insurance is more significant when $\alpha$ is greater. If $0<\alpha<1$, we find that $\frac{\partial^{2} I_t^*}{\partial \alpha^{2}}<0$ when $\alpha<\tilde{\alpha}$ and $\frac{\partial^{2} I_t^*}{\partial \alpha^{2}}>0$ when $\alpha>\tilde{\alpha}$, where $\tilde{\alpha}:=1-\frac{1}{2} \ln(1+\theta_{\epsilon(t)})$. So for low risk-averse investors ($0<\alpha<1$), the magnitude of reduction in insurance depends on the risk aversion attitude.

\subsection{Impact of insurance on value function}
In this subsection, we want to calculate the advantage of buying insurance for investors when facing a random insurable risk.
To achieve this objective, we first assume some investors cannot access the insurance market.
We then calculate the value function with the constraint of no insurance, denoted by $V_1(x,i)$, and compare $V_1(x,i)$ with $V(x,i)$ (the value function of the unconstrained Problem \ref{prob}).

Under the constraint of no insurance, the dynamics of the wealth process $X_1$ is given by
\begin{equation*}
\begin{split}
dX_1(t)&= \left( r_{\epsilon(t)}X_1(t) + (\mu_{\epsilon(t)} - r_{\epsilon(t)}) \pi(t) X_1(t) -c(t) \right) dt\\
&\quad+\sigma_{\epsilon(t)} \pi(t) X_1(t) dW(t) - L_t dN(t), \quad X_1(0)=x.
\end{split}
\end{equation*}
Here the insurable loss $L(t)=\eta_{\epsilon(t)} \, l(t) X_1(t)$.

We then formulate the constrained problem as follows.

\begin{prob} \label{constrained_problem}
Select an admissible policy $u_1^*:=(\pi_1^*,c_1^*)$ that maximizes the criterion function $J$, defined by \eqref{eqn_criterion}. In addition, find the value function
\[V_1(x,i):=\sup_{u_1 \in \mathcal{A}_1} J(x,i;u_1).\]
\end{prob}
For every $u_1=(\pi_1,c_1)\in \mathcal{A}_1$, $\pi_1$ and $c_1$ need to satisfy all the conditions that $\pi$ and $c$  satisfy, where $(\pi,c,I) \in \mathcal{A}_{x,i}$. Since for any $u_1=(\pi_1,c_1)\in \mathcal{A}_1$, we have $(\pi_1,c_1,I\equiv 0) \in \mathcal{A}_{x,i}$. Therefore, $V(x,i) \ge V_1(x,i)$ for all $x>0$ and $i \in \St$.

We provide a verification theorem to Problem \ref{constrained_problem} when the utility function does not depend on the regime (see Theorems \ref{th1} and \ref{th2} for proofs), that is, $U(y,i)=U(y)$ for every $i \in \St$.

\begin{thm} \label{theorem_constrained_problem}
Suppose $U(0)=0$ or $U(0)=-\infty$. Let $v(\cdot,i) \in C^2 (0,\infty) $ be an increasing and concave function such that $v(0,i)=\frac{U(0)}{\delta}$ for every $i \in \St$.
If $v=v(\cdot,\cdot)$ satisfies the Hamilton-Jacobi-Bellman equation
\begin{equation} \label{newhjb1}
\sup_{(\pi_1,c_1)} \big\{ \mathcal{G}_i^{\pi_1,c_1}v(x,i)+U(c_1)+\lambda_i E[v(x-L,i)] \big\} = -\sum_{j\in \St} q_{ij} v(x,j),
\end{equation}
where the operator $\mathcal{G}$ is defined as
\[ \mathcal{G}_i^{\pi_1,c_1} (\psi):=(r_i x +(\mu_i-r_i) \pi x - c_1) \psi' + \frac{1}{2}\sigma_i^2 \pi^2 x^2 \psi'' - (\delta+\lambda_i) \psi,\]
and the control $u_1^*:=(\pi_1^*,c_1^*)$ defined by
\begin{equation*}
u_1^*(t) : = \left( \frac{1}{1-\alpha} \frac{(\mu_{\epsilon(t)}-r_{\epsilon(t)}) }{\sigma_{\epsilon(t)}^2 }, \, (U')^{-1} (v' \left(X_1^*(t), \epsilon(t) \right)) \right)
\end{equation*}
is admissible, then $u_1^*$ is optimal control to Problem \ref{constrained_problem}.
\end{thm}

\subsubsection{$U(y)=\ln(y),\, y>0$}

Under the logarithmic utility, we find the value function to Problem \ref{constrained_problem} is given by
\[ \hat{v}_1(x,i)=\frac{1}{\delta} \ln(\delta x) + \hat{a} _i,\]
where the constants $\hat{a}_i$ satisfy the following linear system
\begin{equation}
\frac{r_i}{\delta} + \frac{\gamma_i}{\delta} + \frac{\lambda_i}{\delta} \hat{\Upsilon}_i  - 1 = \delta \hat{a}_i -\sum_{j \in \St} q_{ij} \hat{a}_j,
\end{equation}
with $\hat{\Upsilon}_i$ defined by $\hat{\Upsilon}_i:=E[\ln(1-\eta_i l)]$.

To compare the value functions $\hat{v}$ and $\hat{v}_1$, we assume there are two regimes ($S=2$) in the economy. Under this assumption, we find $\hat{a}_i$ given by
\begin{equation*}
 \hat{a}_i=\frac{\Pi_i(r_j+\gamma_j-\delta+\lambda_j \hat{\Upsilon}_j)+(\delta + \Pi_j)(r_i+\gamma_i-\delta+\lambda_i \hat{\Upsilon}_i) }{\delta^2(\delta + \Pi_1 + \Pi_2)},
\end{equation*}
where $i,j=1,2$ and $i \neq j$.

We then calculate
\begin{equation} \label{difference_log}
\hat{v}(x,i)-\hat{v}_1(x,i)=\frac{\Pi_i \lambda_j (\hat{\Lambda}_j-\hat{\Upsilon}_j)+\lambda_i(\delta+\Pi_j) (\hat{\Lambda}_i-\hat{\Upsilon}_i)}{\delta^2(\delta + \Pi_1 + \Pi_2)},
\end{equation}
where $i,j=1,2$ and $i \neq j$.

To facilitate our scenario analysis, we assume $\frac{\theta_1}{\eta_1 (1+\theta_1)} \le \frac{\theta_2}{\eta_2 (1+\theta_2)}$ and $l$ is either constant or uniformly distributed on $(0,1)$.

\begin{itemize}
\item Case 1: $l$ is constant.

In this case, $\hat{\Upsilon}_i = \ln(1-\eta_i l)$, $i=1,2$.

(i) Optimal insurance is no insurance for both regimes.

From Example \ref{exm_log}, we notice when the optimal insurance $I^*$ is no insurance,
we have $\hat{\Lambda}_i=\ln(1-\eta_i l)$, $i=1,2$. Then, we obtain $\hat{\Lambda}_i-\hat{\Upsilon}_i=0$ for both regimes. Hence $\hat{v}(x,i)=\hat{v}_1(x,i)$ for all $x>0$ and $i=1,2$.

(ii) Optimal insurance is strictly positive in at least one regime.

When the optimal insurance $I^*$ is strictly positive in at least one regime, we must have at least one $\hat{\Lambda}_i$ in the form of $\hat{\Lambda}_i=-\ln(1+\theta_i)-\eta_il(1+\theta_i)+\theta_i$. Without loss of generality, we assume $I^*>0$ in regime 1, or equivalently, $\eta_1 l - \frac{\theta_1}{1+\theta_1}>0$. Then we obtain
\begin{align*}
\hat{\Lambda}_1-\hat{\Upsilon}_1 &=-\ln(1+\theta_1)-\eta_1l(1+\theta_1)+\theta_1-\ln(1-\eta_1 l)\\
&>-\eta_1 l (1+\theta_1) - \ln(1-\eta_1 l ),
\end{align*}
where the second inequality comes from $-\ln(1+\theta_1)+\theta_1>0$.
Consider $w(l):=-\eta_1 l (1+\theta_1) - \ln(1-\eta_1 l )$. We have $w(0)=0$ and
\[w'(l)= \frac{\eta_1}{(1-\eta_1 l)(1+\theta_1)} \left( \eta_1 l - \frac{\theta_1}{1+\theta_1} \right) >0.\]
This implies $w(l)>0$ for all $l \in (0,1)$, and then $\hat{\Lambda}_1-\hat{\Upsilon}_1>0$. Together with the result above, we can claim that $\hat{\Lambda}_2-\hat{\Upsilon}_2 \ge 0$. Hence, regardless of the optimal insurance $I^*$ in regime 2, we have $\hat{v}(x,i)>\hat{v}_1(x,i)$ for both regimes according to \eqref{difference_log}. Even when $I^*(x,i=2,l)=0$, buying insurance in regime 1 increases the value function in regime 2, which is a surprising result.

To further study the advantage of buying insurance, we define the increase ratio of the value function by
\[ m(x,i):= \left| \frac{V(x,i)-V_1(x,i)}{V_1(x,i)} \right|, \,i=1,2,\]
where $V(x,i)$ and $V_1(x,i)$ are the value functions to Problem \ref{prob} and Problem \ref{constrained_problem}, respectively.

Without loss of generality, we assume $x=\frac{1}{\delta}$ (such assumption makes the constant $\frac{1}{\delta}\ln(\delta x)$ be $0$). Hence, we have $V(x,i)=\hat{v}(x,i)=\hat{A}_i$ and $V_1(x,i)=\hat{v}_1(x,i)=\hat{a}_i$, $i=1,2$. Then we obtain for $i=1,2$
\[ m(x,i)=\frac{\Pi_i \lambda_j (\hat{\Lambda}_j-\hat{\Upsilon}_j)+\lambda_i(\delta+\Pi_j) (\hat{\Lambda}_i-\hat{\Upsilon}_i)}{|\Pi_i(r_j+\gamma_j-\delta+\lambda_j \hat{\Upsilon}_j)+(\delta + \Pi_j)(r_i+\gamma_i-\delta+\lambda_i \hat{\Upsilon}_i)|}. \]
To analyze the impact of the insurable loss on the ratio $m$, we keep $l$ as a variable and choose Parameter Set I but $\delta=0.2$. Notice that for the chosen parameters, our assumption is satisfied
\[\frac{\theta_1}{\eta_1 (1+\theta_1)}=0.16 < \frac{\theta_2}{\eta_2 (1+\theta_2)}=0.2.\]
Since we assume $I^*>0$ in regime $1$, $l \in (0.16,1)$. We draw the graph of the increase ratio of the value function in Figure \ref{pic_compare_valuefunction_log_constant}.
\begin{figure}[ht!]
 \centering
 \includegraphics[width=130mm]{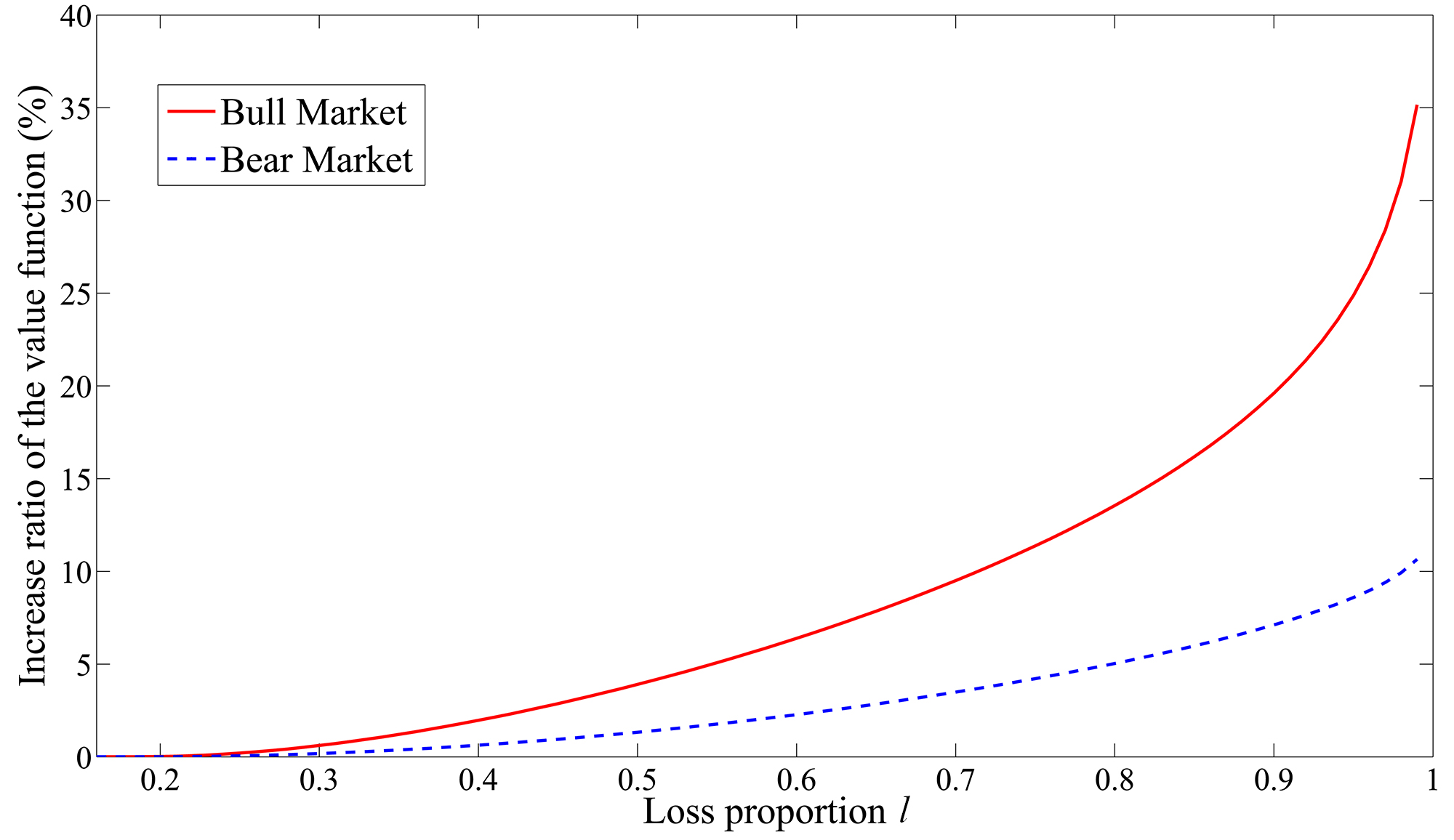}
 \caption{Increase ratio of the value function when loss proportion is constant}
 \label{pic_compare_valuefunction_log_constant}
\end{figure}
As expected, the advantage of buying insurance increases when the insurable loss becomes larger in both regimes. But
surprisingly, we find that buying insurance benefits investors more in a bull market, especially when the insurable loss is large.

\item Case 2. $l$ is uniformly distributed on $(0,1)$.\\
In this case, $\hat{\Upsilon}_i=\int_0^1 \ln(1-\eta_i l) dl = (1-\frac{1}{\eta_i}) \ln(1-\eta_i ) -1$, $i=1,2$.

(i) Optimal insurance is no insurance for both regimes.

In this scenario, it is obvious that $\hat{\Lambda}_i=\hat{\Upsilon}_i$ and then $\hat{v}(x,i)=\hat{v}_1(x,i)$, for all $x>0$ and $i =1,2$.

(ii) Optimal insurance is strictly positive in at least one regime.

Again we assume $I^*>0$ in regime 1. Then we have
\[ \hat{\Lambda}_1-\hat{\Upsilon}_1 = (\frac{1}{\eta_1}-1) \ln((1+\theta_1)(1-\eta_1)) + 1 - \frac{(\eta_1(1+\theta_1) - \theta_1)^2+2\theta_1}{2\eta_1 (1+\theta_1)} .\]
Here $\hat{\Lambda}_1-\hat{\Upsilon}_1$ depends on the premium loading $\theta$ and loss intensity $\eta$ in regime 1. To investigate such dependency, we conduct a numerical calculation. Notice that $\eta_1$ must satisfy the condition $\eta_1 \ge \frac{\theta_1}{1+\theta_1}$.
We draw the difference $\hat{\Lambda}_1-\hat{\Upsilon}_1$ in Figure \ref{difference_log_uniform} when $\theta_1=0.01, 0.1, 0.2, 0.5, 0.8, 0.99$.
\begin{figure}[ht!]
 \centering
 \includegraphics[width=130mm]{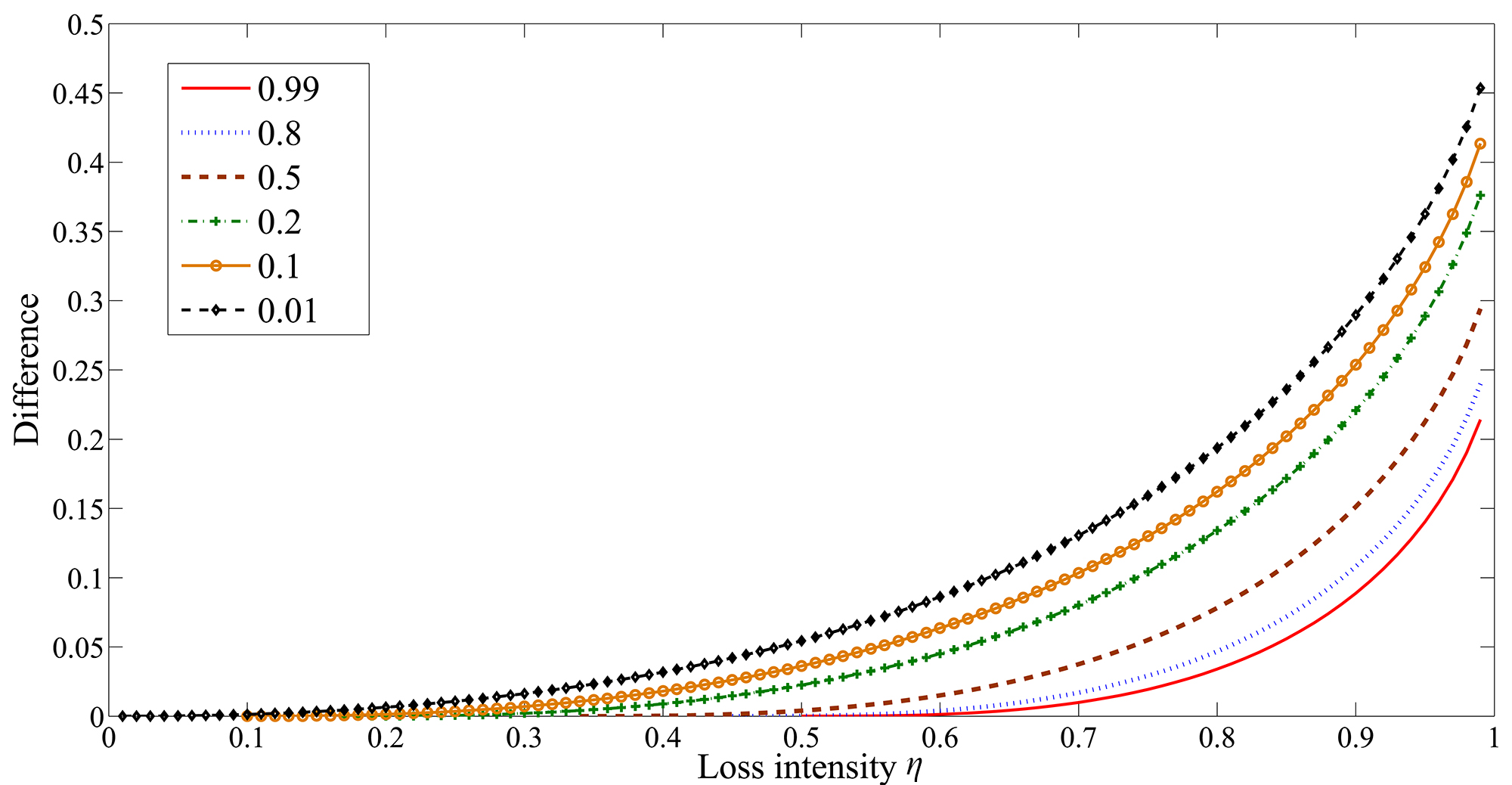}
 \caption{$\hat{\Lambda}_i-\hat{\Upsilon}_i$}
 \label{difference_log_uniform}
\end{figure}
We observe that $\hat{\Lambda}_1-\hat{\Upsilon}_1$ is strictly positive and therefore $\hat{v}(x,i) >\hat{v}_1(x,i)$ for both regimes, which is consistent with our findings in the previous case. Furthermore, as $\theta$ increases (which means the cost of insurance policy increases), the difference of $\hat{\Lambda}_i-\hat{\Upsilon}_i$ becomes smaller, so the benefit of purchasing insurance policy decreases accordingly. Investors gain more advantage from insurance when the insurable loss becomes larger (that is, the loss intensity $\eta$ increases).

\end{itemize}

\subsubsection{$U(y)=-y^\alpha,\,\alpha<0$}
\label{subsubsec_negativepower}

The value function to Problem \ref{constrained_problem} is given by
\[ \tilde{v}_1(x,i)=-\tilde{a}_i ^{1-\alpha} x^\alpha, \]
where positive constants $\tilde{a}_i$  satisfy
\begin{equation} \label{no-insurance1}
 \left(\delta - \alpha r_i - \frac{\alpha}{1-\alpha} \gamma_i + \lambda_i (1-\tilde{\Upsilon}_i)\right) \tilde{a}_i^{1-\alpha} - (1-\alpha) \tilde{a}_i^{-\alpha} = \sum_{j \in \St}
q_{ij} \tilde{a}_j^{1-\alpha},
\end{equation}
with $\tilde{\Upsilon}_i:=E \big[(1-\eta_i l)^\alpha \big]$.

Comparing with the value function we found in Section \ref{subsec_negativepower}, we have $\tilde{v}(x,i) - \tilde{v}_1(x,i) = -(\tilde{A}_i^{1-\alpha} - \tilde{a}_i^{1-\alpha}) x$.

We assume there are two regimes in the economy and the loss proportion $l$ is constant. We skip the trivial case of $I^* \equiv 0$, in which $\tilde{v}(x,i)=\tilde{v}_1(x,i)$ in both regimes. We then carry out a numerical calculation to study the non-trivial case, that is $I^*(x,i,l)>0$ in at least one regime.

To solve the systems \eqref{nonlinear} and \eqref{no-insurance1} numerically, we choose Parameter Set I but $\delta=0.25$. For the chosen parameters, it is more reasonable to consider the case when $l \in (\frac{\nu_2}{\eta_2},1)$ (Since both $\frac{\nu_1}{\eta_1}$ and $\frac{\nu_2}{\eta_2}$ are small).
In Table \ref{table_compare_valuefunction_power} we calculate $\tilde{v}(x,i)-\tilde{v}_1(x,i)$ for various values of $\alpha$  (when calculating $\tilde{v}(x,i)-\tilde{v}_1(x,i)$, we take $x=1$).
\begin{table}[ht!]
\centering
\begin{tabular}{|c|c|c|c|} \hline
$\alpha$ & $l$ &  $\tilde{v}(x,1)-\tilde{v}_1(x,1)$ & $\tilde{v}(x,2)-\tilde{v}_1(x,2)$\\ \hline
& $l=0.30$ & $7.7176 \times 10^{-5}$ & $7.4304 \times 10^{-5}$  \\
-0.01 &  $l=0.60$ & $9.8981 \times 10^{-4}$ & $9.5315 \times 10^{-4}$ \\
& $l=0.90$ & 0.0041 & 0.0039 \\  \hline
& $l=0.15$ &  0.0015 & 0.0015 \\
-0.5 & $l=0.35$ & 0.0797 & 0.0781 \\
& $l=0.50$ & 0.3010 & 0.2961 \\ \hline
& $l=0.20$ & $0.1675 $ & $0.1653$ \\
-1 & $l=0.30$ & 0.8116 & 0.8036 \\
& $l=0.40$ & 2.6969 & 2.6841 \\ \hline
& $l=0.08$ & 0.2454 & 0.2441 \\
-2 & $l=0.10$ & 0.9421 & 0.9381 \\
& $l=0.12$ & 2.2418 & 2.2344 \\ \hline
\end{tabular}
\caption{$\tilde{A}_i-\tilde{a}_i$ when loss proportion $l \in (\frac{\nu_2}{\eta_2}, 1)$}
\label{table_compare_valuefunction_power}
\end{table}
The result clearly confirms that $\tilde{v}(x,i)>\tilde{v}_1(x,i)$ in both regimes. We also observe that the advantage of buying insurance is greater for investors with higher risk aversion. The size of the insurable loss $l$ affects the advantage of buying insurance as well.
When the insurable loss increases (loss proportion $l$ increases), buying insurance will give investors more advantage.
We obtain $\tilde{v}(1,1)-\tilde{v}_1(1,1) > \tilde{v}(1,2)-\tilde{v}_1(1,2)$, meaning buying insurance is more advantageous in a bull market.

\subsubsection{$U(y)=y^\alpha,\, 0<\alpha<1$}
We find the corresponding value function to Problem \ref{constrained_problem} given by
\[\bar{v}_1(x,i)=\bar{a}_i^{1-\alpha} x^\alpha,\]
where the constants $\bar{a}_i$ satisfy the system \eqref{no-insurance1} with $0<\alpha<1$.

From the discussion in Section \ref{subsec_positivepower}, we obtain $\bar{v}(x,i)-\bar{v}_1(x,i)=(\bar{A}_i^{1-\alpha}-\bar{a}_i^{1-\alpha}) x^\alpha$. We then follow all the assumptions made in Section \ref{subsubsec_negativepower} including $x=1$ and conduct a numerical analysis by choosing Parameter Set II.
In this numerical example, we have $\frac{\nu_1}{\eta_1} \le \frac{\nu_2}{\eta_2} < 1$ when $\alpha \in (0,0.8672]$, $\frac{\nu_2}{\eta_2} \le \frac{\nu_1}{\eta_1} < 1$, when $\alpha \in (0.8672,0.9132]$, and $\frac{\nu_2}{\eta_2} \le 1 < \frac{\nu_2}{\eta_2}$ when $\alpha \in (0.9132,1)$.
We consider the first scenario: $\frac{\nu_1}{\eta_1} \le \frac{\nu_2}{\eta_2} < 1$ since it includes most low risk-averse investors. We are interested in the case of $I^*>0$ in at least one regime. For the chosen parameters, we find $\frac{\nu_1}{\eta_1}$ is so small that the case of $l \in (0,\frac{\nu_1}{\eta_1})$ is rare.
So we further assume constant loss proportion $l \in (\frac{\nu_1}{\eta_1} , \frac{\nu_2}{\eta_2}]$. Notice that when $l \in (\frac{\nu_1}{\eta_1} , \frac{\nu_2}{\eta_2}]$, we have $I^*(x,1,l)>0$ but $I^*(x,2,l)=0$.

\begin{figure}[ht!]
 \centering
 \includegraphics[width=130mm]{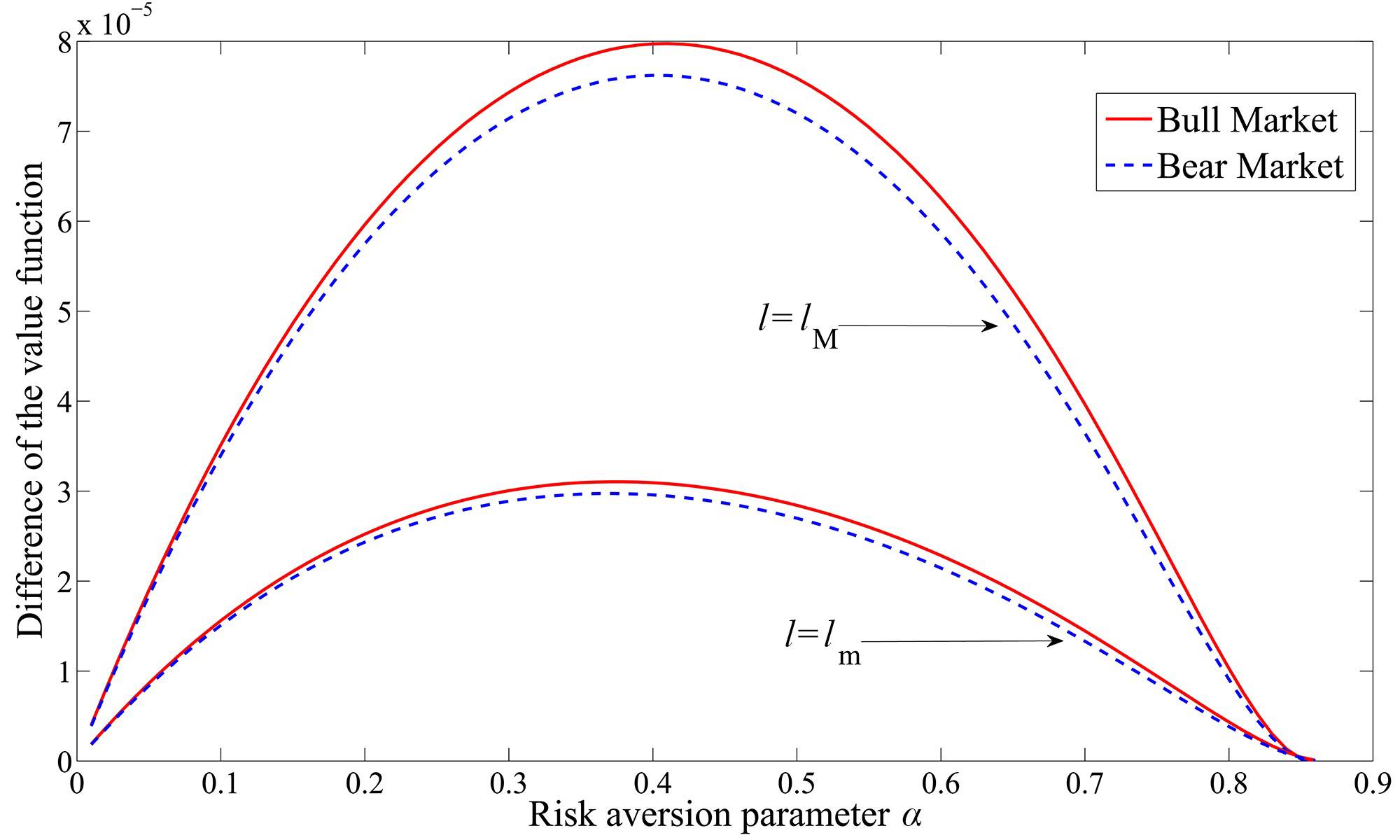}
 \caption{$\bar{v}(1,i)-\bar{v}_1(1,i)$ when $l \in (\frac{\nu_1}{\eta_1} , \frac{\nu_2}{\eta_2}]$}
 \label{pic_compare_valuefunction_power_constant}
\end{figure}

From solving the non-linear systems \eqref{nonlinear1} and \eqref{no-insurance1}, we draw $\bar{v}(x,i)-\bar{v}_1(x,i)$ for $l=l_M:=\frac{1}{2}(\frac{\nu_1}{\eta_1}+\frac{\nu_2}{\eta_2})$ and $l=l_m:=\frac{\nu_2}{\eta_2}-0.01$ in Figure \ref{pic_compare_valuefunction_power_constant}.
It is obvious that $\bar{v}(1,i)-\bar{v}_1(1,i)>0$ in both regimes. As have seen in the previous cases, the benefit of buying insurance in a bull market strictly outperforms that in a bear market. We also observe a surprising result that the difference of the value functions (advantage of buying insurance) is not an increasing function of $\alpha$, which is different from the result in Section \ref{subsubsec_negativepower}.  But the difference is a concave function of $\alpha$.

\section{Conclusions}
\label{sec_conclusion}
We have considered simultaneous optimal consumption, investment and insurance problems in a regime switching model which enables the regime of the economy to affect not only the financial  but also the insurance market. A risk-averse investor facing an insurable risk wants to obtain the
optimal consumption, investment and insurance policy that maximizes her/his expected total discounted utility of consumption over an infinite time horizon.

We have presented the first versions of verification theorems for simultaneous optimal consumption, investment and insurance problems when there is regime switching. We have also obtained explicitly  the optimal policy and the value function when the utility function belongs to the HARA class.

The optimal proportion of wealth invested in the stock is constant in every regime, and is greater in a bull market regardless of the investor's risk aversion attitude. We observe that
investors with high risk tolerance invest a large proportion of wealth in the stock.

The optimal consumption to wealth ratio is a strictly increasing function of  the investor's risk aversion parameter ($\alpha$).
Moderate risk-averse investors ($\alpha=0$) consume at a constant proportion in both regimes.
High risk-averse investors ($\alpha<0$) consume a higher proportion of their wealth in a bull market.
In contrast, low risk-averse investors ($0<\alpha<1$) consume proportionally more in a bear market.

The optimal insurance is proportional to the investor's wealth and such proportion depends on the premium loading $\theta$ and the investor's risk aversion parameter $\alpha$. As the loading $\theta$ increases, the demand for insurance decreases. This decrease of the demand for insurance is more significant when $\theta$ is small. We observe that investors who are very risk tolerant (that is, investors with large $\alpha$) spend a small amount of wealth in insurance. For high and moderate risk-averse investors ($\alpha \le 0$), the amount of reduction in insurance is greater when $\alpha$ is
far away from 0. However, low risk-averse investors ($0<\alpha<1$) reduce the amount of insurance in different magnitudes that depend on the value of $\alpha$.

We have also obtained the conditions under which it is optimal to buy insurance, and
analyzed their dependence on the different parameters.

We have calculated the advantage of buying insurance.
Based on a comparative analysis, we find the value function $V(x,i)$ to Problem \ref{prob} is strictly greater than the value function $V_1(x,i)$ to Problem \ref{constrained_problem} when the optimal insurance is not equal to $0$ in all regimes.
We also observe that the advantage of buying insurance is greater in a bull market. Investors who face a large random loss,  gain more benefits from purchasing insurance.

\section*{Acknowledgements}
The work of B. Zou and A. Cadenillas was supported by the Natural Science and Engineering Research Council of Canada. The results of this paper were presented at the 2nd Industrial-Academic Workshop on Optimization in Finance and Risk Management, Fields Institute for Research in Mathematical Sciences.

\bibliographystyle{model2-names}

\end{document}